%% file: QCD-10-007_temp.tex
\pdfoutput=1

\documentclass[11pt,twoside,a4paper,cmspaper,final,collab]{cms-tdr}
\def\svnVersion{57931}\def\cmsCernNo{2010-094}\def\cmsCernDate{\today}\def\cmsMessage{Submitted to the Journal of High Energy Physics}

\begin{document}\cmsNoteHeader{QCD-10-007}

\hyphenation{had-ron-i-za-tion}
\hyphenation{cal-or-i-me-ter}
\hyphenation{de-vices}
\RCS$Revision: 55412 $
\RCS$HeadURL: svn+ssh://alverson@svn.cern.ch/reps/tdr2/papers/QCD-10-007/trunk/QCD-10-007.tex $
\RCS$Id: QCD-10-007.tex 55412 2011-05-13 10:19:49Z stenson $
\cmsNoteHeader{QCD-10-007} 
\title{
Strange Particle Production in pp Collisions at $\sqrt{s}$ = 0.9 and 7~TeV}

\ifthenelse{\boolean{cms@external}}{%
\author{Kevin Stenson}\affiliation{University of Colorado, Boulder}
}
{%
\address[cu]{University of Colorado, Boulder}
\author[cu]{Kevin Stenson}
}

\date{\today}

\abstract{

  The spectra of strange hadrons are measured in proton-proton
  collisions, recorded by the CMS experiment at the CERN LHC, at
  centre-of-mass energies of 0.9 and 7\,TeV\@.  The
  $\mathrm{K}_\mathrm{S}^0$, $\Lambda$, and $\Xi^-$ particles and
  their antiparticles are reconstructed from their decay topologies
  and the production rates are measured as functions of rapidity and
  transverse momentum, \pt.
The results are
  compared to other experiments and to predictions of the \PYTHIA Monte
  Carlo program.  The \pt distributions are found to differ
  substantially from the \PYTHIA results and the production rates
  exceed the predictions by up to a factor of three.}

\hypersetup{
pdfauthor={CMS Collaboration},%
pdftitle={Strange Particle Production in pp Collisions at sqrt(s) = 0.9 and 7 TeV},%
pdfsubject={CMS},%
pdfkeywords={CMS, physics, QCD, strange, production}
}

\maketitle 

\section{Introduction}

Measurements of particle yields and spectra are an essential step in understanding
proton-proton collisions at the Large Hadron Collider (LHC).  The Compact Muon Solenoid
(CMS) Collaboration has published results
on spectra of charged particles at centre-of-mass energies of 0.9, 2.36, and
7~TeV\,\cite{CMS_QCD-09-010,CMS_QCD-10-006}. In this analysis the measurement is
extended to strange mesons and baryons ($\mathrm{K}_\mathrm{S}^0$, $\Lambda$,
$\Xi^-$)\,\footnote{Particle-conjugate states are implied throughout this paper.}
at centre-of-mass energies of 0.9 and 7~TeV\@.
The investigation of strange hadron production is an important ingredient in understanding the nature
of the strong force.  The LHC experiments ALICE and LHCb have recently reported results on strange hadron
production at $\sqrt{s} = 0.9$~TeV~\cite{AliceStrange,Aaij:2010nx}.  In addition to results at
$\sqrt{s} = 0.9$\,TeV, we also present results at $\sqrt{s} = 7$\,TeV, opening up a new
energy regime in which to study the strong interaction.
As the strange quark is heavier than
up and down quarks, production of strange hadrons is generally suppressed relative to
hadrons containing only up and down quarks.  The amount of strangeness suppression is
an important component in Monte Carlo (MC) models such as \PYTHIA\,\cite{Pythia}
and \textsc{hijing/b}$\overline{\hbox{\sc b}}$\,\cite{Pop:2010qz}.  Because the threshold for strange
quark production in a quark-gluon plasma is much smaller than in a hadron gas, an
enhancement in strange particle production has frequently been suggested as an
indication of quark-gluon plasma formation\,\cite{Koch:1986ud}.  This effect would be further
enhanced in baryons with multiple strange quarks.
While a quark-gluon plasma is more likely to be found in collisions of heavy nuclei,
the enhancement of strange quark production in high energy pp collisions would be a sign of
a collective effect, according to some models\,\cite{Abreu:2007kv,Werner:2010zz}.  In contrast,
recent Regge-theory calculations indicate little change in the ratio of $\mathrm{K}_\mathrm{S}^0$
to charge particle production with increasing collision
energy\,\cite{Likhoded:2010pc,Chliapnikov:1988xa}.
Thus, these measurements can be used to constrain theories, provide input
for tuning of Monte Carlo models, and serve as a reference for the interpretation of
strangeness production results in heavy-ion collisions.

Minimum bias collisions at the LHC can be classified as elastic scattering, inelastic
single-diffractive dissociation (SD), inelastic double-diffractive dissociation, and
inelastic non-diffractive scattering. The results presented here are normalized to the sum
of double-diffractive and non-diffractive interactions, referred to as
non-single-diffractive (NSD) interactions\,\cite{CMS_QCD-09-010,CMS_QCD-10-006}.  This
choice is made to most closely match the event selection and to compare with previous
experiments, which often used similar criteria.  The $\mathrm{K}_\mathrm{S}^0$, $\Lambda$,
and $\Xi^-$ are long-lived particles ($c \tau > 1~\mathrm{cm}$) and can be identified from
their decay products originating from a displaced vertex. The particles are reconstructed
from their decays: $\mathrm{K}_\mathrm{S}^0 \rightarrow \pi^+ \pi^-$, $\Lambda\rightarrow
\mathrm{p} \pi^-$, and $\Xi^- \rightarrow \Lambda \pi^-$ over the rapidity range $|y|<2$, where the rapidity
is defined as $y=\frac{1}{2}\ln{\frac{E+p_L}{E-p_L}}$, $E$ is the particle energy, and $p_L$ is the
particle momentum along the anticlockwise beam direction.  For each particle species, we
measure the production rate versus rapidity and transverse momentum
$\pt$, the average $\pt$, the central production rate $\frac{dN}{dy}|_{y\approx 0}$,
and the integrated yield for $|y|<2$ per NSD event. We compare our measurements to results from
Monte Carlo models and lower energy data.

\section{CMS experiment and collected data}
\label{sec:cms}

CMS is a general purpose experiment at the LHC\,\cite{CMS_Detector}.
The silicon tracker, lead-tungstate crystal electromagnetic calorimeter, and brass-scintillator
hadron calorimeter are all immersed in a 3.8\,T axial magnetic field while muon detectors
are interspersed with flux return steel outside of the 6~m diameter superconducting solenoid.
The silicon tracker is used to reconstruct charged particle trajectories with
$|\eta|<2.5$, where the pseudorapidity is defined as $\eta = -\ln{\tan{\frac{\theta}{2}}}$,
$\theta$ being the polar angle with respect to the anticlockwise beam.
The tracker consists of layers of $100\!\times\!150\;\mu\mathrm{m}^2$ pixel sensors at
radii less than 15~cm and layers of strip sensors, with pitch ranging from 80 to 183~$\mu$m,
covering radii from 25 to 110~cm.  In addition to barrel and endcap detectors,
CMS has extensive forward calorimetry including a steel and quartz-fibre hadron
calorimeter (HF), which covers $2.9 < |\eta| < 5.2$.
The data presented in this paper were collected by the CMS experiment
in spring 2010 from proton-proton collisions at centre-of-mass energies of 0.9 and 7\TeV
during a period in which the probability for two collisions in the same bunch crossing
was negligible and the bunch crossings were well separated.

The online selection of events required activity in the beam scintillator counters at
$3.23 < |\eta| < 4.65$ in coincidence with colliding proton bunches.  The offline selection
required deposits of at least 3\GeV of energy in each end of the
HF\,\cite{CMS_QCD-09-010}, preferentially selecting NSD events. A
primary vertex reconstructed in the tracker was required and beam-halo and other
beam-related background events were rejected as described in
Ref.\,\cite{CMS_QCD-09-010}.  The data selected with these criteria contain
9.08 and 23.86 million events at 0.9 and 7~TeV, corresponding
to approximate integrated luminosities of 240 and 480~$\mu$b$^{-1}$,
respectively.
To determine the acceptance and efficiency, minimum-bias Monte Carlo
samples were generated at both centre-of-mass energies using \PYTHIA
6.422\,\cite{Pythia} with tune D6T\,\cite{Bartalini:2010su}.  These events were passed through a
CMS detector simulation package based on \GEANT4\,\cite{Geant4}.

\section{Strange particle reconstruction}

Ionization deposits recorded by the silicon tracker are used to reconstruct tracks.
To maximize reconstruction efficiency, we use a combined track collection formed from
merging tracks found with the standard tracking described in Ref.\,\cite{CMS_TRK-10-001} and the
minimum bias tracking described in Ref.\,\cite{CMS_QCD-09-010}.  Both tracking collections
use the same basic algorithm; the differences are in the requirements for seeding,
propagating, and filtering tracks.

As described in Ref.\,\cite{CMS_TRK-10-001}, the $\mathrm{K}_\mathrm{S}^0$ and $\Lambda$
(generically referred to as $\mathrm{V}^0$) reconstruction combines pairs of oppositely charged tracks;
if the normalized $\chi^2$ of the fit to a common vertex is less than 7, the candidate is kept.
The primary vertex is refit for each candidate, removing the two tracks associated with the
$\mathrm{V}^0$ candidate.
The next two paragraphs describe the selection of candidates for measurement of $\mathrm{V}^0$
and $\Xi^-$ properties, respectively.  Selection variables are measured in units of $\sigma$,
the calculated uncertainty including all correlations.

\begin{figure}[hbtp]
  \begin{center}
    \includegraphics[width=0.45\textwidth]{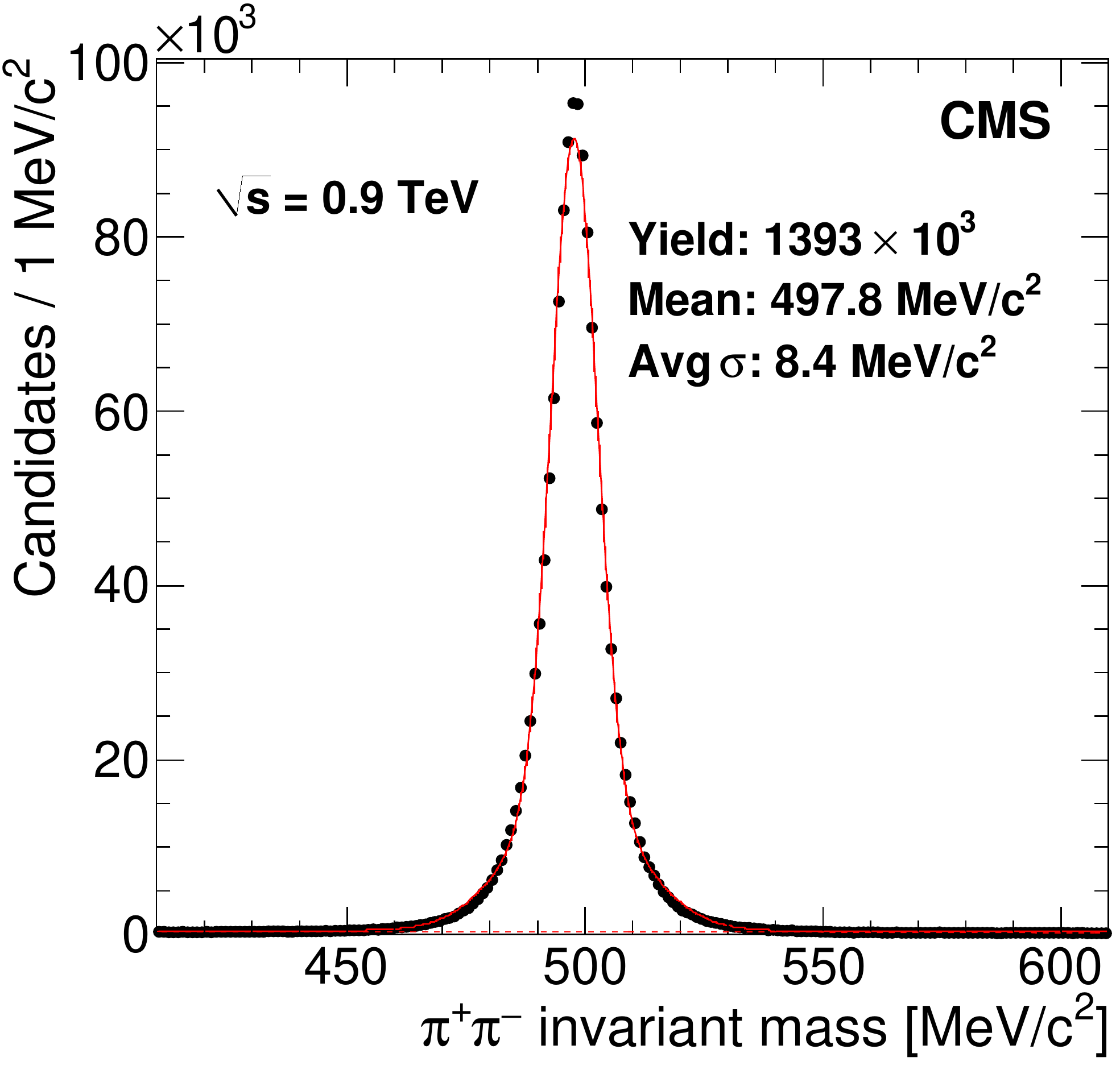}\hspace{5pt}
    \includegraphics[width=0.45\textwidth]{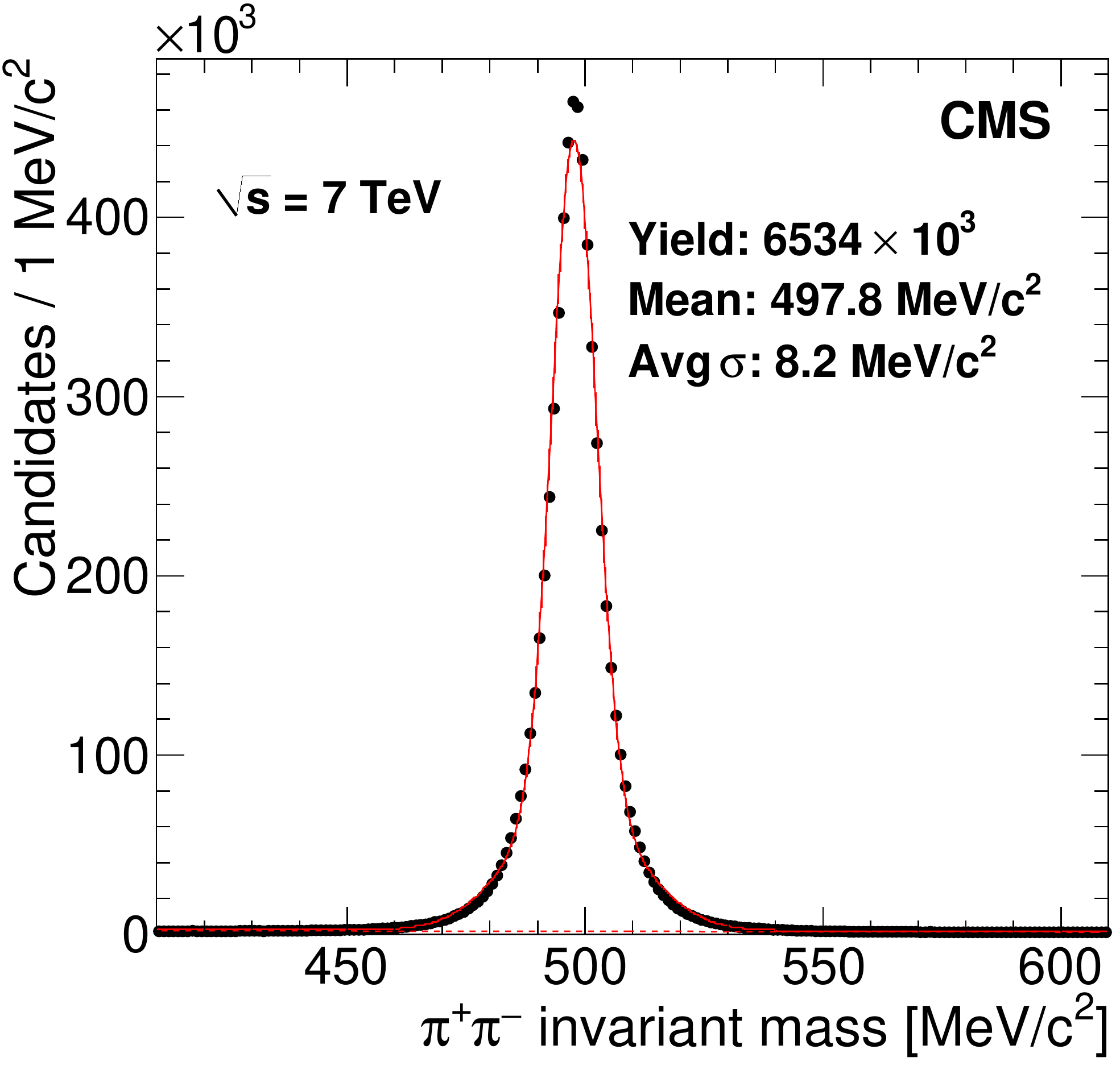}
    \caption{The $\pi^+\pi^-$ invariant mass distributions from data collected at $\sqrt{s} = 0.9\TeV$ (left) and
7\TeV (right).  The solid curves are fits to a double Gaussian and quadratic polynomial.
The dashed curves show the quadratic background contribution.}
    \label{fig:kshort_yield}
  \end{center}
\end{figure}

\begin{figure}[hbtp]
  \begin{center}
    \includegraphics[width=0.45\textwidth]{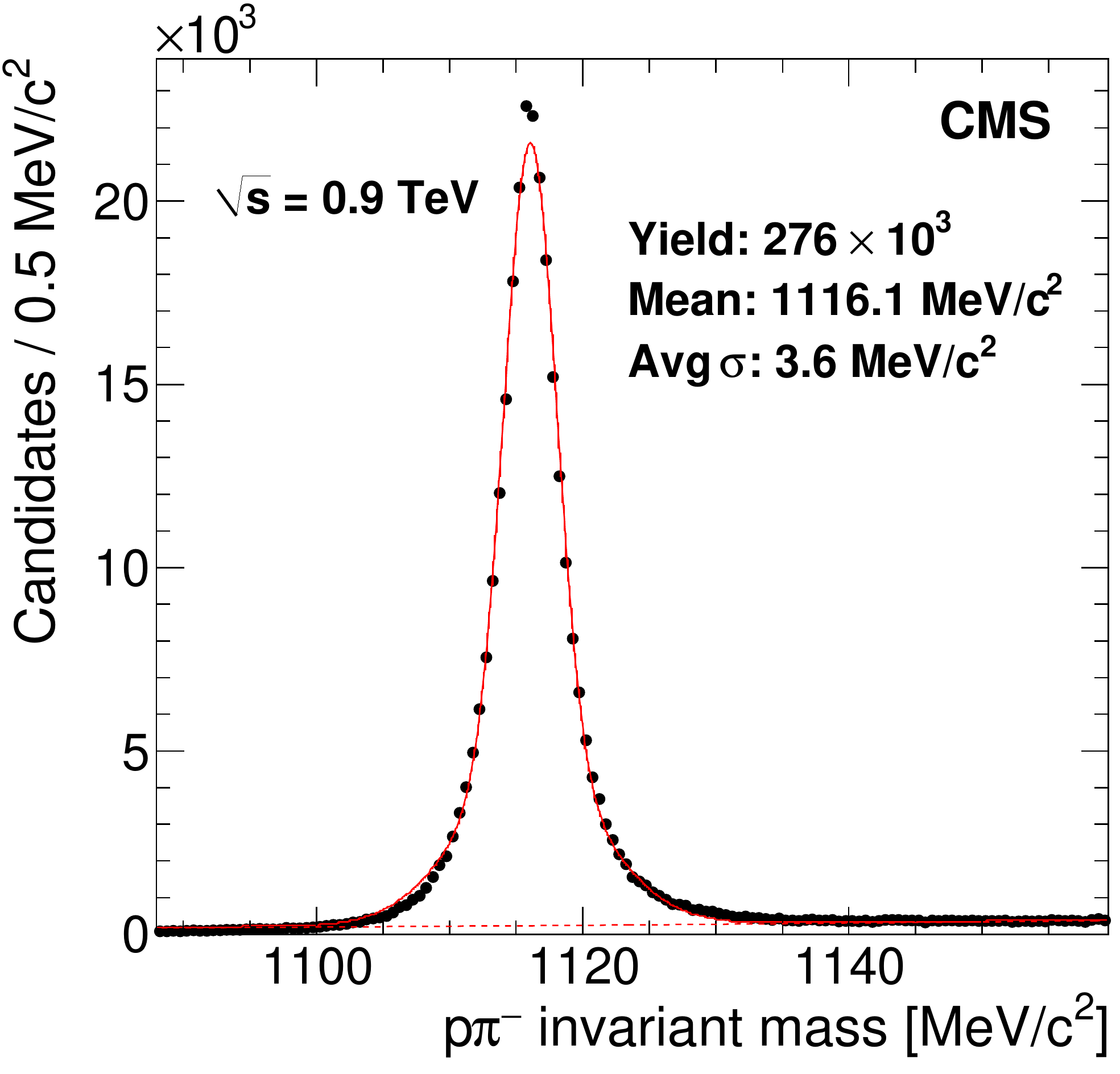}\hspace{5pt}
    \includegraphics[width=0.45\textwidth]{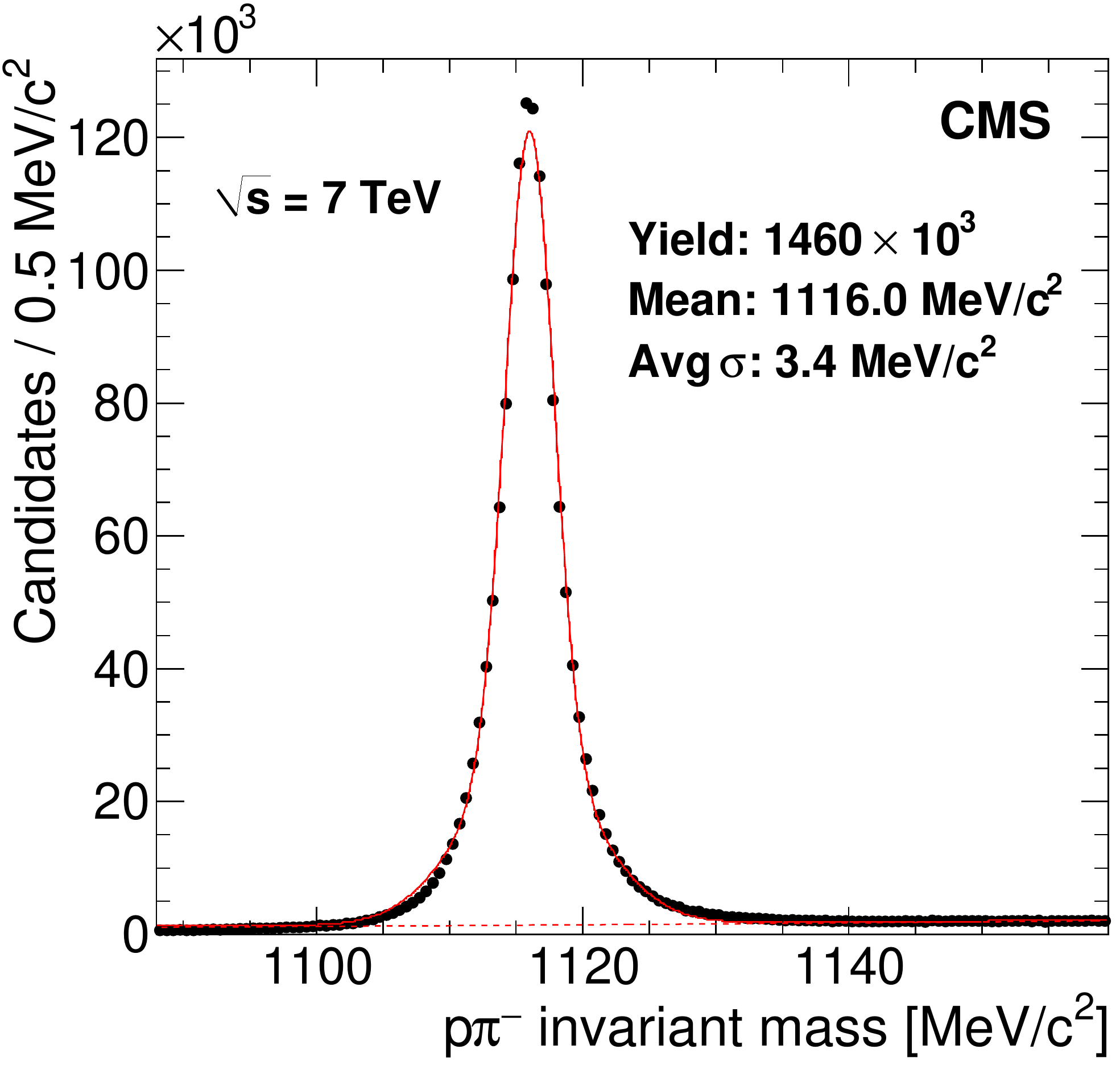}
    \caption{The $p\pi^-$ invariant mass distributions from data collected at $\sqrt{s} = 0.9\TeV$ (left) and
7\TeV (right).  The solid curves are fits to a double Gaussian signal and a background function
given by $Aq^{B}$, where $q=M_{p\pi^-}-(m_p+m_{\pi^-})$.  The dashed curves show the background contribution.}
    \label{fig:lambda_yield}
  \end{center}
\end{figure}

To remove $\mathrm{K}_\mathrm{S}^0$ particles misidentified as $\Lambda$ particles and vice versa,
the $\mathrm{K}_\mathrm{S}^0 (\Lambda)$ candidates must have
a corresponding $p\pi^- (\pi^+\pi^-)$ mass more than $2.5\sigma$ away from the
world-average $\Lambda (\mathrm{K}_\mathrm{S}^0)$ mass.
The production cross sections we measure are intended to represent the prompt
production of $\mathrm{K}_\mathrm{S}^0$ and $\Lambda$, including strong and electromagnetic decays.
However, $\mathrm{V}^0$ particles can also be produced from weak decays and from secondary nuclear interactions.
These unwanted contributions are reduced by requiring that the $\mathrm{V}^0$ momentum vector points back to
the primary vertex.  This is done by requiring the 3D distance of closest approach of the
$\mathrm{V}^0$ to the primary vertex to be less than 3$\sigma$.  To remove generic prompt backgrounds,
the 3D $\mathrm{V}^0$ vertex separation from the primary vertex must be greater than 5$\sigma$ and
both $\mathrm{V}^0$ daughter tracks must have a 3D distance of closest approach to the primary
vertex greater than 3$\sigma$.
With the above selection, the background level for low transverse-momentum $\Lambda$ candidates remains
high.  Therefore, additional cuts are applied to $\Lambda$ candidates with $\pt < 0.6\,\mathrm{GeV}/c$:
\begin{itemize}
\item{3D separation between the primary and $\Lambda$ vertices $>10\sigma$ (instead of $>5\sigma$),}
\item{transverse (2D) separation between the pp collision region (beamspot) and $\Lambda$ vertex $>10\sigma$ 
(instead of no cut), where the uncertainty is dominated by the $\Lambda$ vertex, and}
\item{3D impact parameter of the pion and proton tracks with respect to the primary vertex $>(7-2|y|)\sigma$
(instead of $>3\sigma$) where $y$ is the rapidity of the $\Lambda$ candidate.
The rapidity dependence is a consequence of the observation that, for the low transverse momentum
candidates, large backgrounds dominate at small rapidity, while low efficiency characterizes the large rapidity behaviour.}
\end{itemize}
The resulting mass distributions of $\mathrm{K}_\mathrm{S}^0$ and $\Lambda$ candidates
from the 0.9 and 7~TeV data are shown in Figs.~\ref{fig:kshort_yield} and
\ref{fig:lambda_yield}.  The $\pi^+\pi^-$ mass distribution is fit with a double Gaussian
(with a common mean) signal function plus a quadratic background.  The $p\pi^-$ mass
distribution is fit with a double Gaussian (common mean) signal function and a background
function of the form $Aq^{B}$, where $q = M_{p\pi^-}-(m_p + m_{\pi^-})$, $M_{p\pi^-}$ is
the $p\pi^-$ invariant mass, and $A$ and $B$ are free parameters.  The fitted
$\mathrm{K}_\mathrm{S}^0\: (\Lambda)$ yields at $\sqrt{s} = $ 0.9 and 7~TeV are $1.4\!\times\!
10^6\: (2.8 \!\times\! 10^5)$ and $6.5 \!\times\! 10^6\: (1.5 \!\times\! 10^6)$, respectively.

\begin{figure}[hbtp]
  \begin{center}
    \includegraphics[width=0.45\textwidth]{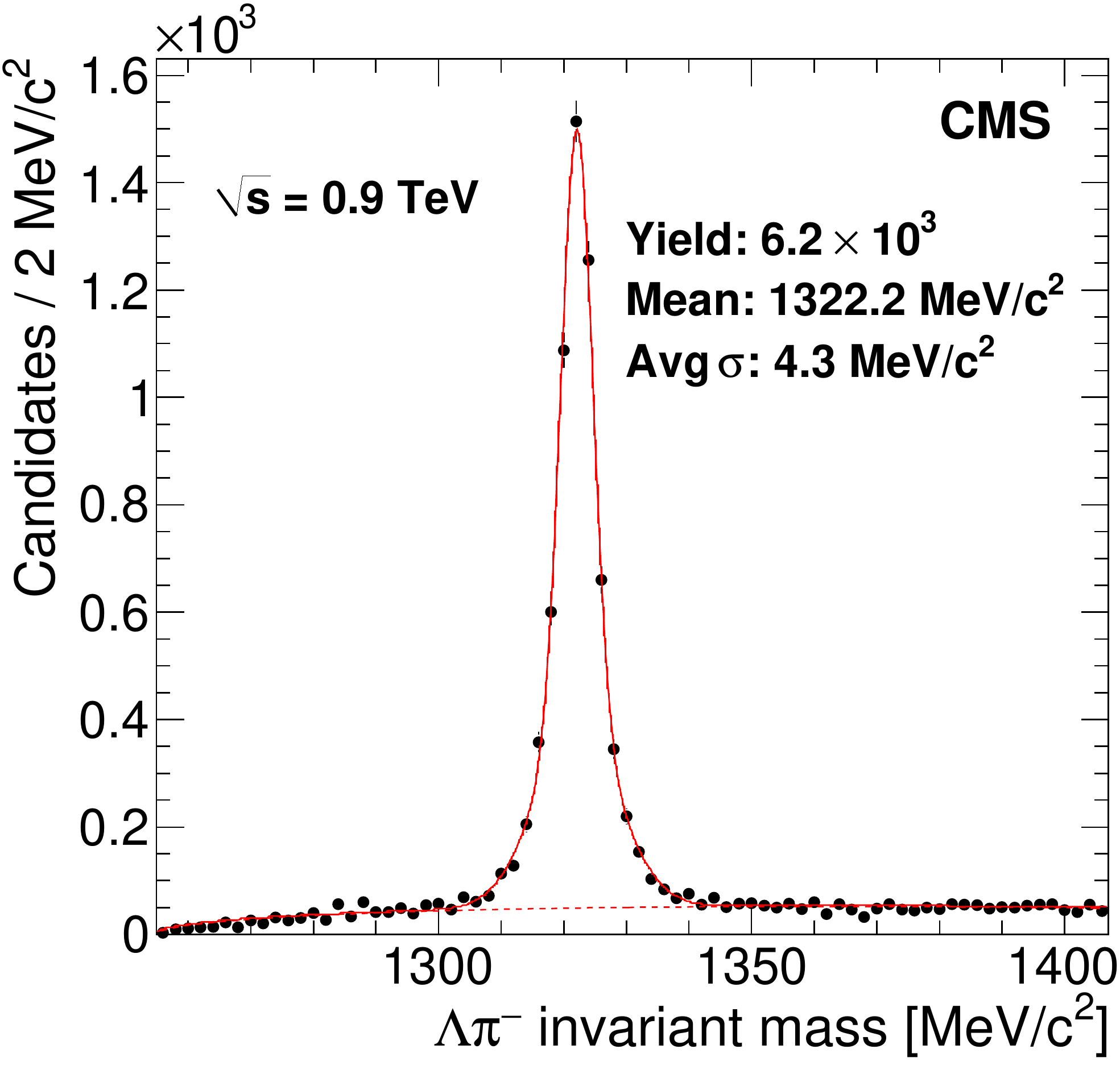}\hspace{5pt}
    \includegraphics[width=0.45\textwidth]{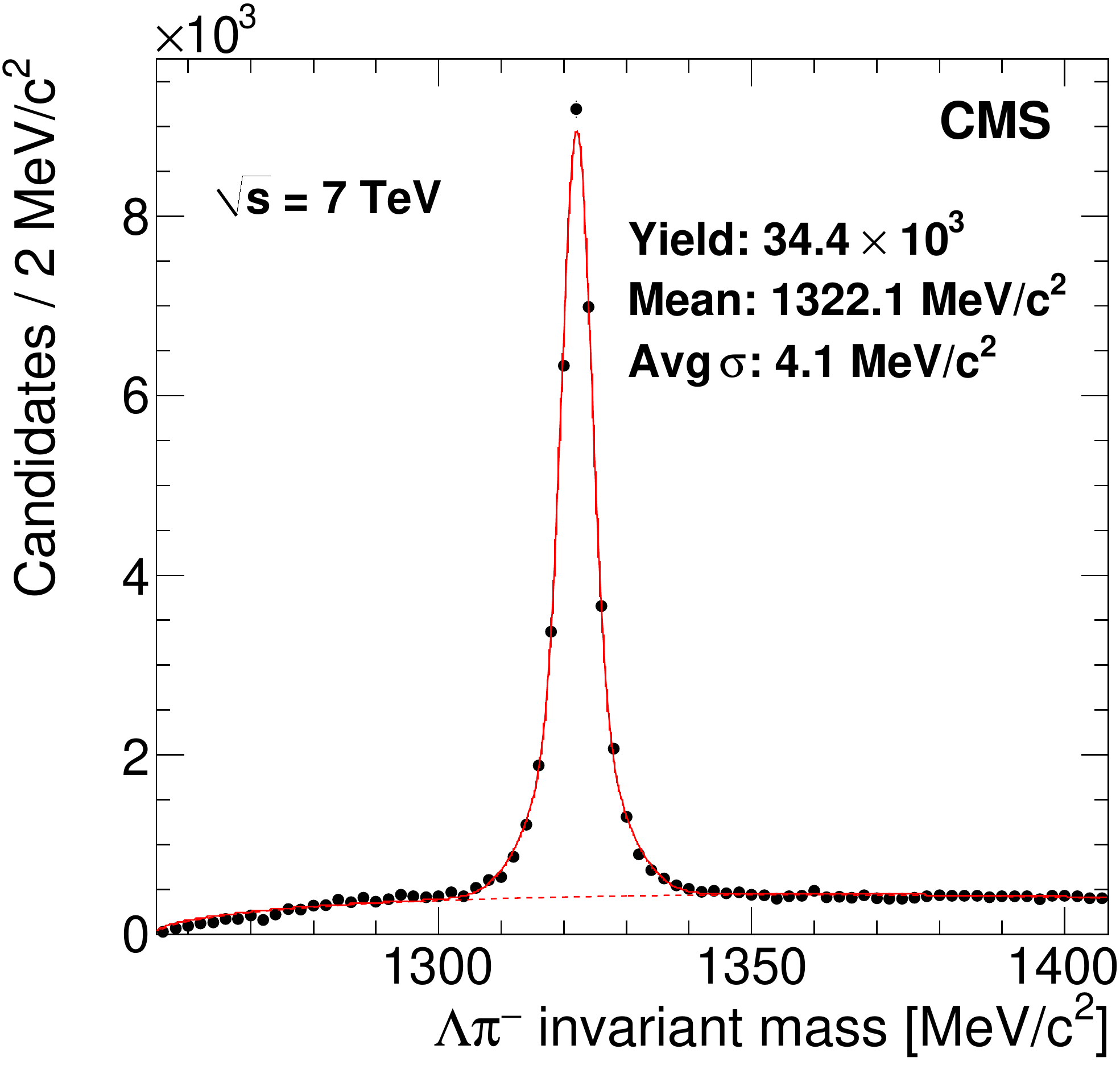}
    \caption{The $\Lambda\pi^-$ invariant mass distributions from data collected at $\sqrt{s} = 0.9\TeV$ (left)
7\TeV (right).  The solid curves are fits to a double Gaussian signal and a background function
given by $Aq^{1/2}+Bq^{3/2}$, where $q=M_{\Lambda\pi^-}-(m_\Lambda+m_{\pi^-})$.  The dashed curves show the background contribution.}
    \label{fig:xi_yield}
  \end{center}
\end{figure}

To reconstruct the $\Xi^-$, charged tracks of the correct sign are combined with $\Lambda$
candidates.  The $\chi^2$ probability of the fit to a common vertex for the $\Lambda$ and the charged
track must be greater than 5\%.  In this fit, the $\Lambda$ candidate is constrained to have the correct
world-average mass\,\cite{PDG}.  The primary vertex is refit for each $\Xi^-$ candidate, removing all
tracks associated with the $\Xi^-$.  The $\Xi^-$ candidates must then pass the following selection criteria:
\begin{itemize}
\item{3D impact parameter with respect to the primary vertex $>2\sigma$ for the proton track from the
$\Lambda$ decay, $>3\sigma$ for the $\pi^-$ track from the $\Lambda$ decay, and $>4\sigma$ for the $\pi^-$ track
from the $\Xi^-$ decay,}
\item{invariant mass from the $\pi^+\pi^-$ hypothesis for the tracks associated with the $\Lambda$ candidate
at least $20\,\mathrm{MeV}/c^2$ away from the world-average $\mathrm{K}_\mathrm{S}^0$ mass,}
\item{3D impact parameter of the $\Xi^-$ candidate with respect to the primary vertex $<3\sigma$,}
\item{3D separation between $\Lambda$ vertex and primary vertex $>10\sigma$, and}
\item{3D separation between $\Xi^-$ vertex and primary vertex $>2\sigma$.}
\end{itemize}
The mass distributions of $\Xi^-$ candidates from the $\sqrt{s} =$ 0.9 and 7~TeV data are shown in
Fig.~\ref{fig:xi_yield}.  The $\Lambda\pi^-$ mass is fit with a double Gaussian (with a common mean)
signal function and a background function of the form $Aq^{1/2}+Bq^{3/2}$, where
$q = M_{\Lambda\pi^-}-(m_\Lambda + m_{\pi^-})$ and $M_{\Lambda\pi^-}$ is the
$\Lambda\pi^-$ invariant mass.  The fitted $\Xi^-$ yields at $\sqrt{s} = $ 0.9 and 7~TeV are
$6.2\!\times\! 10^3$ and $3.4 \!\times\! 10^4$, respectively.

\section{Efficiency correction}
\label{sec:eff}

The efficiency correction is determined from a Monte Carlo simulation which is
used to measure the effects of acceptance and the efficiency for event selection (including the trigger) and
particle reconstruction.  The Monte Carlo samples are reweighted to match the observed
track multiplicity in data, as this has been shown to be an important component of the
trigger efficiency\,\cite{CMS_QCD-09-010,CMS_QCD-10-006}.  This is referred to as track
weighting. The efficiency correction also accounts for the other decay channels
of the strange particles that we do not attempt to reconstruct, such as $\mathrm{K}_\mathrm{S}^0 \to \pi^0\pi^0$.

The efficiency is given by the number of reconstructed particles divided by
the number of generated particles, subject to two modifications.  Firstly, the efficiency
correction is used to account for candidates from SD events.  As the results are normalized
to NSD events, candidates from SD events which pass the event selection must be removed.  This is done by defining
the efficiency as the number of reconstructed candidates in all events divided by the number of
generated candidates in NSD events.  Secondly, the efficiency is modified to account for the small
contribution of reconstructed non-prompt strange particles which pass the selection
criteria.  This is only an issue for the $\Lambda$ particles which receive contributions
from $\Xi$ and $\Omega$ decays.  Since these non-prompt $\Lambda$ particles are present in
both the MC and data, we modify the efficiency to remove this contribution by calculating the
numerator using all of the reconstructed strange particles and the denominator with only the prompt generated
strange particles.  As the MC fails to produce enough $\Xi$ particles (see Section~\ref{sec:results}), the
non-prompt $\Lambda$'s are weighted more than prompt $\Lambda$'s in the efficiency calculation.

The results of this analysis are presented in terms of two kinematic distributions:
transverse momentum and rapidity.  For all modes, $|y|$ is divided into 10 equal size bins
from 0 to 2 and \pt is divided into 20 equal size bins from 0 to 4\GeVc plus one bin each
from 4 to 5\GeVc and 5 to 6\GeVc.  In addition, the $\mathrm{V}^0$ modes also have 6--8\GeVc and
8--10\GeVc \pt bins.  All results are for particles with $|y|<2$.

\begin{figure}[hbtp]
  \begin{center}
    \includegraphics[width=0.48\textwidth]{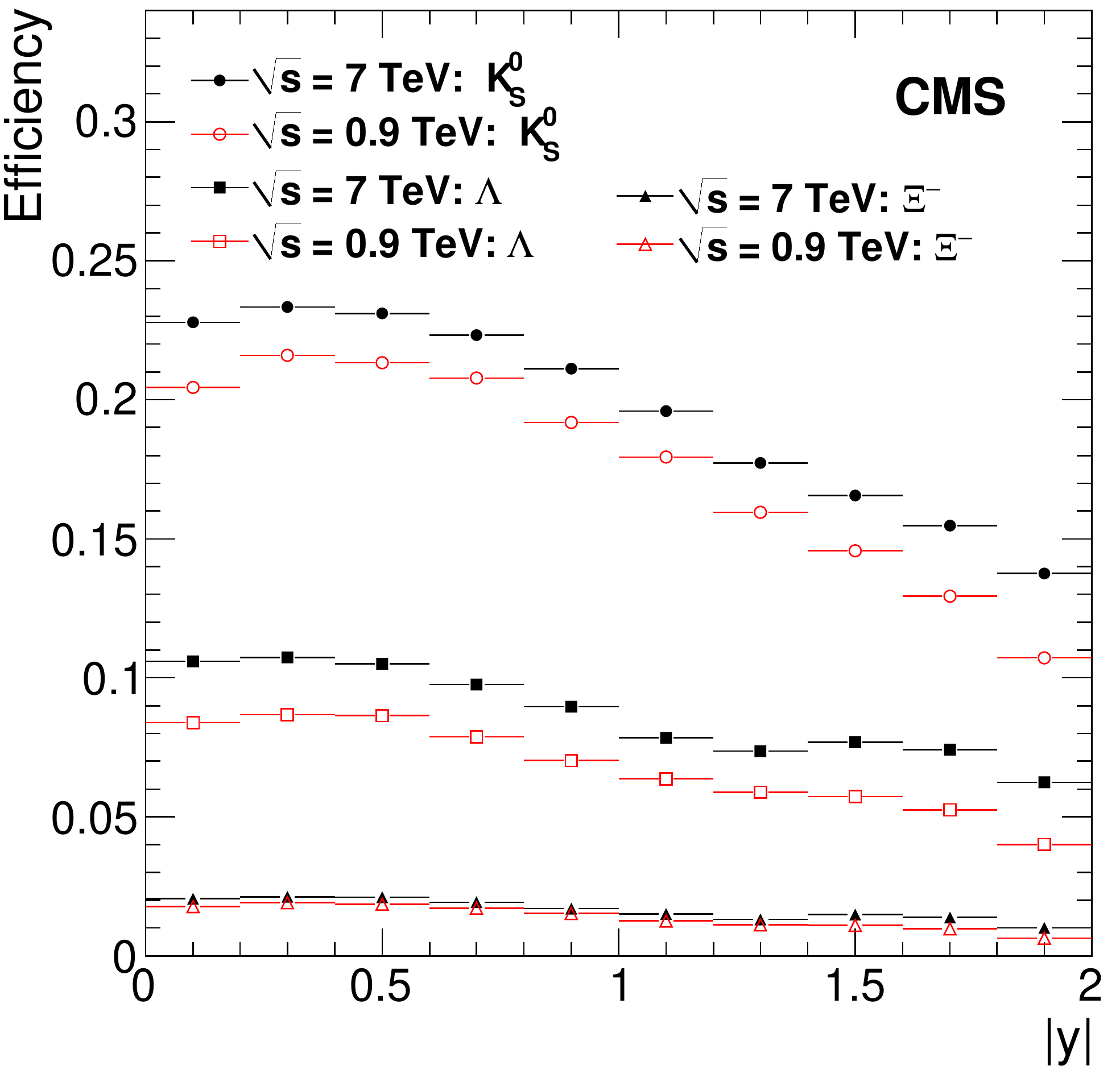}
    \includegraphics[width=0.48\textwidth]{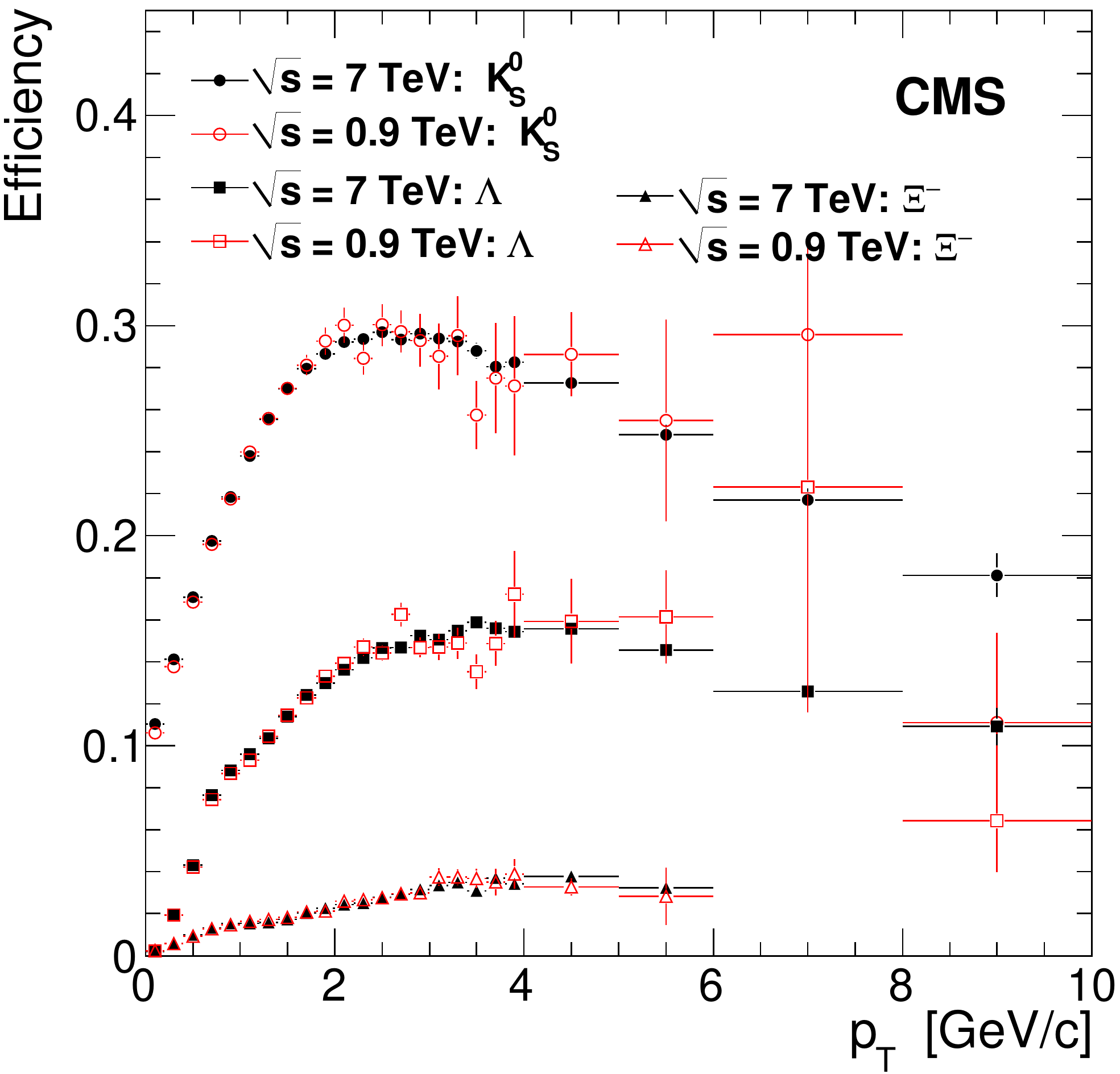}
    \caption{Total efficiencies, including acceptance, trigger and event selection,
reconstruction and particle selection, and other decay modes, as a function
of $|y|$ (left) and $\pt$ (right) for $\mathrm{K}_\mathrm{S}^0$, $\Lambda$, and $\Xi^-$ produced
promptly in the range $|y|<2$. Error bars come from MC statistics.}
    \label{fig:efficiency}
  \end{center}
\end{figure}

The efficiency correction for the $\mathrm{V}^0$ modes uses a two-dimensional binning in \pt and
$|y|$.  Thus, the data are divided into 240 bins in the $|y|,\pt$ plane.  The invariant mass histograms
in each bin are fit to a double Gaussian signal function (with a common mean) and a background
function.  In bins with few entries, a single Gaussian signal function is used.  For
the $\Lambda$ sample, some bins are merged due to sparse populations in $|y|,\pt$
space.  The merging is performed separately when measuring $|y|$ and $\pt$ such that the
merging occurs across $\pt$ and $|y|$ bins, respectively.  The efficiency from MC is
evaluated in each bin and applied to the measured yield to obtain the corrected yield.
The two-dimensional binning used for the $\mathrm{V}^0$ efficiency correction greatly reduces
problems arising from remaining differences in production dynamics between the data and the
simulation.  The much smaller sample of $\Xi^-$ candidates prevents the use of 2D binning.
Thus, the data are divided into $|y|$ bins to measure the $|y|$
distribution and into $\pt$ bins to measure the $\pt$ distribution.  However,
the MC spectra do not match the data.  Therefore, each Monte Carlo $\Xi^-$
particle is weighted in \pt ($|y|$) to match the distribution in data when measuring the
efficiency versus $|y|$ (\pt). Thus, the MC and data distributions are forced to match in
the variable over which we integrate to determine the efficiency.  We refer to this as
kinematic weighting.  The efficiencies for all three particles are shown versus $|y|$ and
\pt in Fig.~\ref{fig:efficiency}.  The efficiencies (for particles with $|y|<2$) include
the acceptance, event selection, reconstruction and selection, and
also account for other decay channels.  The increase in efficiency with $\pt$ is
due to the improvement in tracking efficiency as track $\pt$ increases and to the selection
criteria designed to remove prompt decays.  The slight decrease at high $\pt$ is due to
particles decaying too far out to have reconstructed tracks.  While there is no centre-of-mass
energy dependence on the efficiency versus $\pt$, particles produced at
$\sqrt{s} = 7$\,TeV have a higher average-$\pt$, resulting in a higher efficiency
when plotted versus rapidity.

\begin{figure}[htbp]
  \begin{center}
    \includegraphics[width=0.329\textwidth]{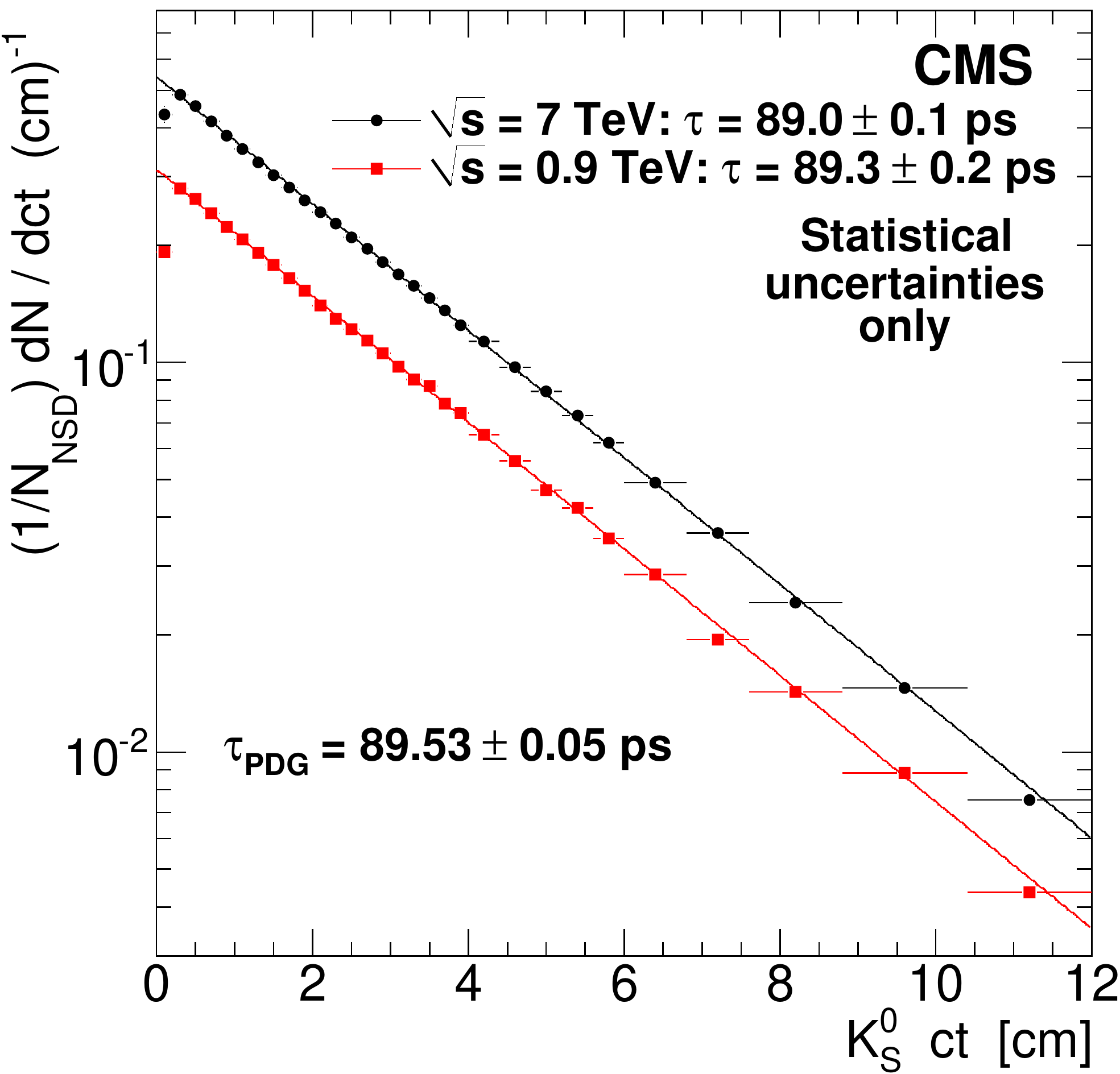}
    \includegraphics[width=0.329\textwidth]{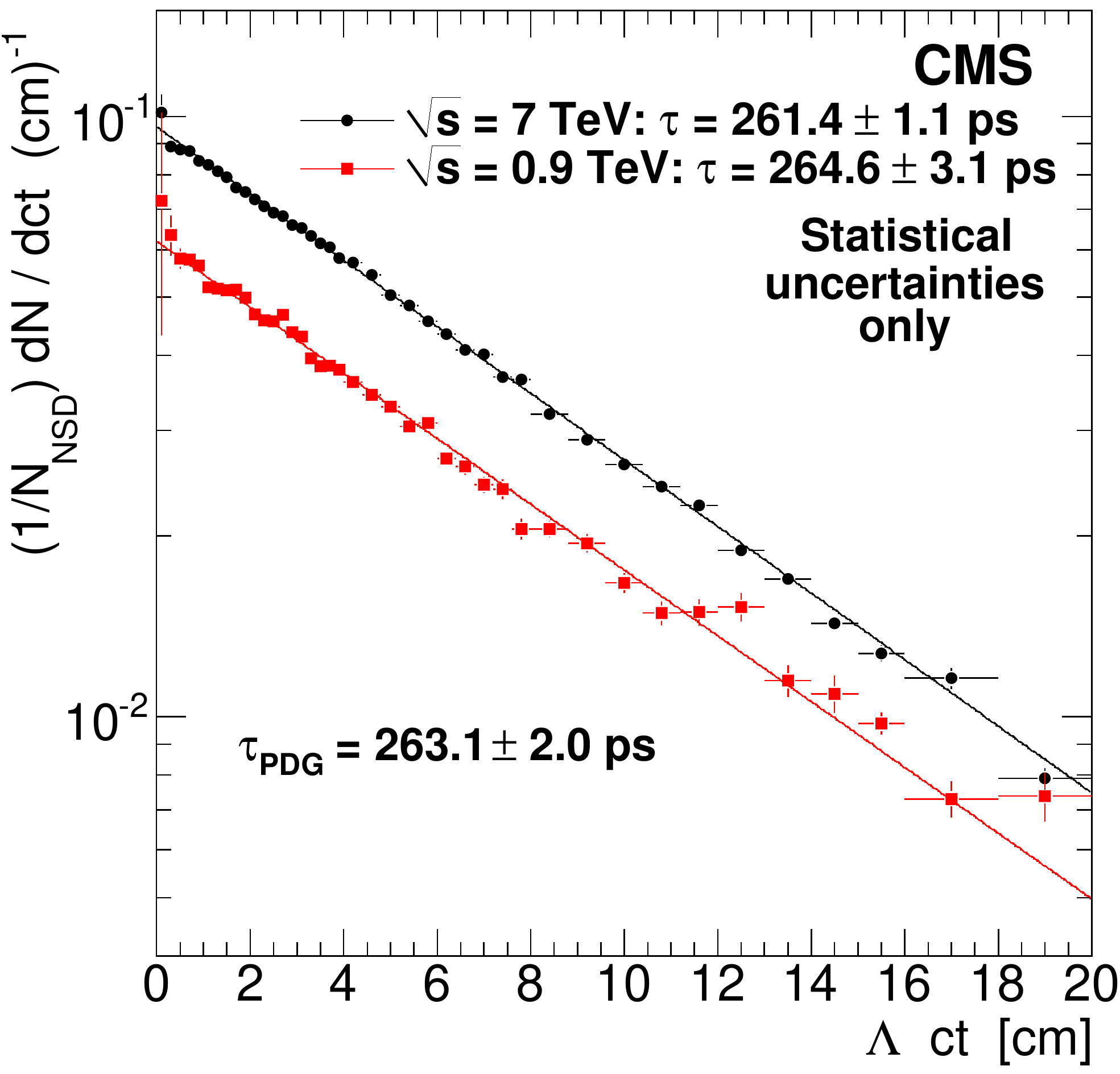}
    \includegraphics[width=0.329\textwidth]{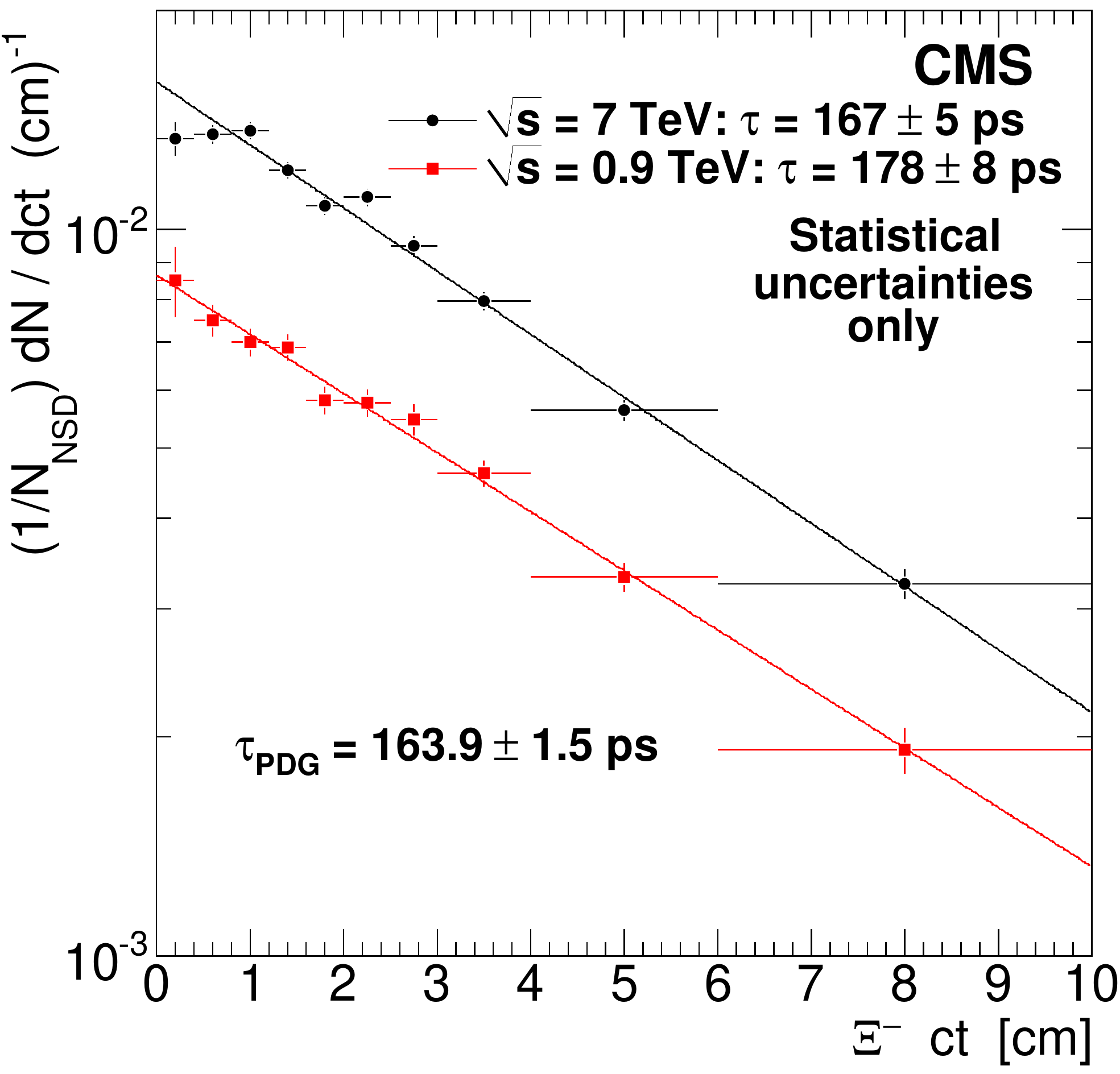}
    \caption{$\mathrm{K}_\mathrm{S}^0$ (left), $\Lambda$ (middle), and $\Xi^-$ (right) corrected
decay time distributions at $\sqrt{s} =$ 0.9 and 7~TeV\@.
The values of the lifetimes, derived from a fit
with an exponential function (solid line), are shown in the legend along with the world-average value.
The error bars and uncertainties on the lifetimes refer to the statistical uncertainty only.}
    \label{fig:lifetime}
  \end{center}
\end{figure}

As a check on the ability of the Monte Carlo simulation to reproduce the efficiency, the
(well-known) $\mathrm{K}_\mathrm{S}^0$, $\Lambda$, and $\Xi^-$ lifetimes are measured.
For the $\mathrm{K}_\mathrm{S}^0$ measurement, the data are divided into bins of \pt and
$ct$, where $ct$ is calculated as $ct = cmL/p$ where $m$, $L$, and $p$ are, respectively, the
mass, decay length, and momentum of the particle.  In each bin the data is corrected by
the MC efficiency and the corrected yields summed in \pt to obtain the $ct$ distribution.
Due to smaller sample sizes, the $\Lambda$ and $\Xi^-$ yields are only measured in bins of $ct$.
Using the kinematic weighting technique, the MC efficiency in each bin of $ct$ is
calculated with the \pt spectrum correctly
weighted to match data.  The corrected lifetime distributions, shown in
Fig.~\ref{fig:lifetime}, display exponential behaviour.  The vertex separation requirements
result in very low efficiencies and low yields in the first lifetime bin and are thus
expected to have some discrepancies.  An actual measurement of the lifetime would remove this
issue by using the reduced proper time, where one measures the lifetime relative to the point at which the particle had a chance
to be reconstructed.  The measured values of the lifetimes are also
reasonably consistent with the world averages\,\cite{PDG} (shown in
Fig.~\ref{fig:lifetime}) considering that only statistical uncertainties are reported and
that this is not the optimal method for a lifetime measurement.

To convert the efficiency corrected yields to per event yields requires the true
number of NSD events, which is obtained by correcting the number of selected events for the
event selection inefficiency.  The event selection includes both the online trigger and offline
selection described in Section~\ref{sec:cms}.  The event selection efficiency is determined in
two ways.  In the default method, it is calculated directly from the
Monte Carlo simulation (appropriately weighted by the track multiplicity to reproduce the
data).  In the alternative method, the event selection efficiency versus track multiplicity is
derived from the Monte Carlo.  Then, each measured event is weighted by the inverse of the
event selection efficiency based on its number of tracks.  The number of
events divided by the number of weighted events gives the event selection
efficiency. However, since the event selection requires a primary vertex, no events will have
fewer than two tracks.  Therefore, the Monte Carlo is also used to determine the fraction
of NSD events which have fewer than two tracks and the event selection efficiency is adjusted to
include this effect.  In both methods, the event selection efficiency accounts for unwanted SD
events which pass the event selection.  The numerator in the efficiency ratio contains all
selected events, including single-diffractive events, while the denominator contains all
NSD events.

\section{Systematic uncertainties}
\label{sec:syst}

The systematic uncertainties, reported in Table~\ref{tab:syst}, are divided into two categories:
normalization uncertainties, which only affect the overall normalization, and point-to-point
uncertainties, which may also affect the shape of the $\pt$ and $|y|$ distributions.

The list below summarizes the source and evaluation of the point-to-point systematic uncertainties.
\begin{itemize}
\item \underline{Kinematic weighting versus 2D binning}: The efficiency corrections using
the 1D kinematic weighting technique (used for the $\Xi^-$ analysis) and the 2D binning
technique (used for the $\mathrm{V}^0$ analysis) were compared by measuring the efficiency with
both methods on the highest statistics channel ($\mathrm{K}_\mathrm{S}^0$ at 7\TeV).
\item \underline{Non-prompt $\Lambda$}: The contribution of non-prompt $\Lambda$ decays is varied
by a factor of two in the simulation.
\item \underline{MC tune}: The nominal efficiency calculated from the default
\PYTHIA6 D6T tune\,\cite{Bartalini:2010su} is compared to the efficiency obtained from the
\PYTHIA6 Perugia0 (P0) tune\,\cite{Skands:2009zm} and \PYTHIA8\,\cite{Pythia8}.
\item \underline{Variation of reconstruction cuts}: The following cuts are varied for
all three modes: $\mathrm{V}^0$ vertex separation significance ($\pm 2\sigma$), 3D impact parameter
of $\mathrm{V}^0$ and $\Xi^-$ ($\pm 2\sigma$), 3D impact parameter of tracks ($\pm 2\sigma$), cut
on $\mathrm{K}_\mathrm{S}^0 (\Lambda)$ mass for $\Lambda (\mathrm{K}_\mathrm{S}^0)$
candidates ($\pm 1.5\sigma$), and increase of number of hits required on each track from 3
to 5.  For the $\Xi^-$, additional cuts were varied: the $\Xi^-$ vertex separation
significance ($\pm 1\sigma$) and $\Xi^-$ vertex fit probability $(\pm 3\%)$.
\item \underline{Detached particle reconstruction}: Finding that the corrected lifetime
distributions are exponential with the correct lifetime is a verification of our
understanding of the reconstruction efficiency versus decay length.
The systematic uncertainty is taken as the difference between the fitted lifetimes
and the world-average lifetimes\,\cite{PDG}.  While the $\mathrm{K}_\mathrm{S}^0$ and
$\Lambda$ lifetimes are within 1\% of the world-average, a 2\% systematic uncertainty is
conservatively assigned.
\item \underline{Mass fits}: As an alternative to using a double-Gaussian signal shape,
the $\mathrm{V}^0$ invariant mass distributions are fit using a signal shape taken from Monte Carlo.
\item \underline{Matching versus fitting}: The number of reconstructed events,
used in the numerator of the efficiency, is calculated in two ways.  The truth matching
method counts all reconstructed candidates which are matched to a generated candidate,
based on the daughter momentum vectors and the decay vertex.  The fitting method fits the
MC mass distributions to extract a yield.  The difference between these two is taken as a
systematic.
\item \underline{Misalignment}: The nominal efficiency, obtained using a realistic
alignment in the MC, is compared to the efficiency from a MC sample with perfect
alignment.
\item \underline{Beamspot}: The location and width of the luminous region of pp collisions
(beamspot) is varied in the simulation to assess the effect on efficiency.
\item \underline{Detector material}: The nominal efficiency is compared to the efficiency from a
MC simulation in which the tracker was modified.  The modification consisted of two parts.
First, the mass of the tracker was increased by 5\% which is a conservative estimate of
the uncertainty.  Second, the amounts of the various materials inside the tracker were
adjusted within estimated uncertainties to obtain the tracker which maximized the
interaction cross section.  Both effects were implemented by changing material densities
such that the tracker geometry remained the same.  The effect is to decrease the
efficiency as more particles, both primary and secondary, interact.
\item \underline{\GEANT4 cross sections}: The cross sections used by \GEANT4 for low energy strange
baryons and all antibaryons are known to be overestimated\,\cite{Alice_ppbar}.  The size
of this effect is evaluated by analyzing $\Lambda$--$\overline{\Lambda}$ asymmetries.
\end{itemize}

As the trigger efficiency is used to derive the number of NSD events, it only affects the
normalization.  The normalization systematic uncertainties, most of which come from
trigger efficiency uncertainties, are described below.

\begin{itemize}
\item \underline{Alternative trigger efficiency calculation}: The difference between the
default and alternative trigger efficiency measurements, described in Sec.~\ref{sec:eff},
is taken as the systematic uncertainty on the method.
\item \underline{Fraction of SD vs NSD}:  The change in trigger efficiency when the
fraction of single-diffractive events in Monte Carlo is varied by $\pm$50\% is taken as the
systematic uncertainty on the fraction of SD events.  The \PYTHIA6 MC produces approximately
20\% SD events while the fraction in the triggered data is considerably
less\,\cite{CMS_QCD-09-010,CMS_QCD-10-006}.  As the UA5
experiment measured 15.5\% for this fraction at 900\GeV\,\cite{Ansorge:1986xq}, a variation
of $\pm$50\% is conservative.
\item \underline{Modelling diffractive events}: In addition to the fraction of SD events,
the modelling of SD and NSD events may not be correct.  The trigger efficiency obtained
using the D6T tune is compared with the trigger efficiency from the P0 tune and \PYTHIA8.
In particular, \PYTHIA8 uses a new Pomeron description of diffraction,
modelled after PHOJET\,\cite{Engel:1995sb,Bopp:1998rc},
which results in a large increase in the track multiplicity of SD events.
\item \underline{Track weighting}: The track weighting of the Monte Carlo primarily affects
the trigger efficiency.  The track weighting requires a measurement of the track
multiplicity distribution in data and MC\@.  The default track multiplicity distribution
is calculated from events which pass the trigger, except the primary vertex requirement is
not applied.  Two variations are considered. First, the track multiplicity distribution is
measured from events also requiring a primary vertex.  As this requires at least two
tracks per event, the weight for events with fewer than two tracks is taken to be the same
as the weight for events with two tracks.  Second, the track weighting is determined with
the primary vertex requirement (as in the first case), but without the HF trigger.  The
variation is taken as a systematic uncertainty on the track weighting.
\item \underline{Branching fractions}: The results are corrected for other decay channels of
$\mathrm{K}_\mathrm{S}^0$, $\Lambda$, and $\Xi^-$.  The branching fraction uncertainty
reported by the PDG\,\cite{PDG} is used as the systematic uncertainty.
\end{itemize}

The systematic uncertainties at the two centre-of-mass energies are found to be essentially the same.
The normalization uncertainties and the detached particle reconstruction uncertainty are obtained
from the average of the results from the two centre-of-mass energies.
The other point-to-point systematic uncertainties are derived from the higher statistics
7\TeV results.  The point-to-point systematic uncertainties are measured as functions of $\pt$
and $|y|$ and found to be independent of both variables.  Therefore, the systematic
uncertainties are estimated such that they include approximately 68\% of the points (representing
a 1$\sigma$ error).  The resulting systematic uncertainties are summarized in Table~\ref{tab:syst}.
\begin{table}[htbp]
\begin{center}
\caption{Systematic uncertainties for the $\mathrm{K}_\mathrm{S}^0$, $\Lambda$,
and $\Xi^-$ production measurements.}
\begin{tabular}{lcccccc}
\hline
\hline
Source & $\mathrm{K}_\mathrm{S}^0$ (\%) & $\Lambda$ (\%) & $\Xi^-$ (\%) \\
\hline
Point-to-point systematic uncertainties   \\
~~~~Kinematic weight vs. 2D binning & 1.0 & 1.0 & 1.0  \\
~~~~Non-prompt $\Lambda$ & --- & 3.0 & --- \\
~~~~MC tune                & 2.0 & 3.0 & 4.0 \\
~~~~Reconstruction cuts    & 4.0 & 5.0 & 5.0 \\
~~~~Detached particle reconstruction  & 2.0 & 2.0 & 3.5 \\
~~~~Mass fits              & 0.5 & 2.0 & 2.0 \\
~~~~Matching vs. fitting   & 2.0 & 3.0 & 3.0 \\
~~~~Misalignment           & 1.0 & 1.0 & 1.0 \\
~~~~Beamspot               & 1.0 & 1.5 & 2.0 \\
~~~~Detector material      & 2.0 & 5.0 & 8.0 \\
~~~~\GEANT 4 cross sections& 0.0 & 5.0 & 5.0 \\
\hline
Point-to-point sum   & 5.9 & 10.7 & 12.7 \\
\hline
\hline
Normalization systematic uncertainties  \\
~~~~Trigger calculation     & 1.8 & 1.8 & 1.8 \\
~~~~SD fraction             & 2.8 & 2.8 & 2.8 \\
~~~~Diffractive modelling    & 1.5 & 1.5 & 1.5 \\
~~~~Track weighting         & 2.0 & 2.0 & 2.0 \\
~~~~Branching fractions        & 0.1 & 0.8 & 0.8 \\
\hline
Normalization sum & 4.1 & 4.2 & 4.2 \\
\hline
\hline
Overall sum  & 7.2 & 11.5 & 13.4 \\
\hline
\hline
\end{tabular}
\label{tab:syst}
\end{center}
\end{table}

For the measurements of $dN/dy$, $dN/dy|_{y\approx 0}$, and $dN/d\pt$, the full systematic
uncertainty is applied.  For the $\Lambda/\mathrm{K}_\mathrm{S}^0$ and $\Xi^-/\Lambda$
production ratio measurements, the largest point-to-point systematic uncertainty
of the two particles is used and, among the normalization systematic uncertainties, only
the branching fraction correction is considered.  Note that for the $\Xi^-/\Lambda$
production ratios, the $\Lambda$ branching fraction uncertainty cancels in the ratio.

\section{Results}
\label{sec:results}

The results reported here are normalized to NSD interactions. The number of NSD
raw events (given in Sec.~\ref{sec:cms}) are corrected for the trigger efficiency
and the fraction of SD events after the selection.
The corrected number of NSD events is
$9.95\!\times\! 10^6$ and $37.10\!\times\! 10^6$ for $\sqrt{s}=$ 0.9 and 7~TeV, respectively.

\subsection{Distributions $dN/dy$ and $dN/d\pt$}
\begin{figure}[hbtp]
  \begin{center}
    \includegraphics[width=0.42\textwidth]{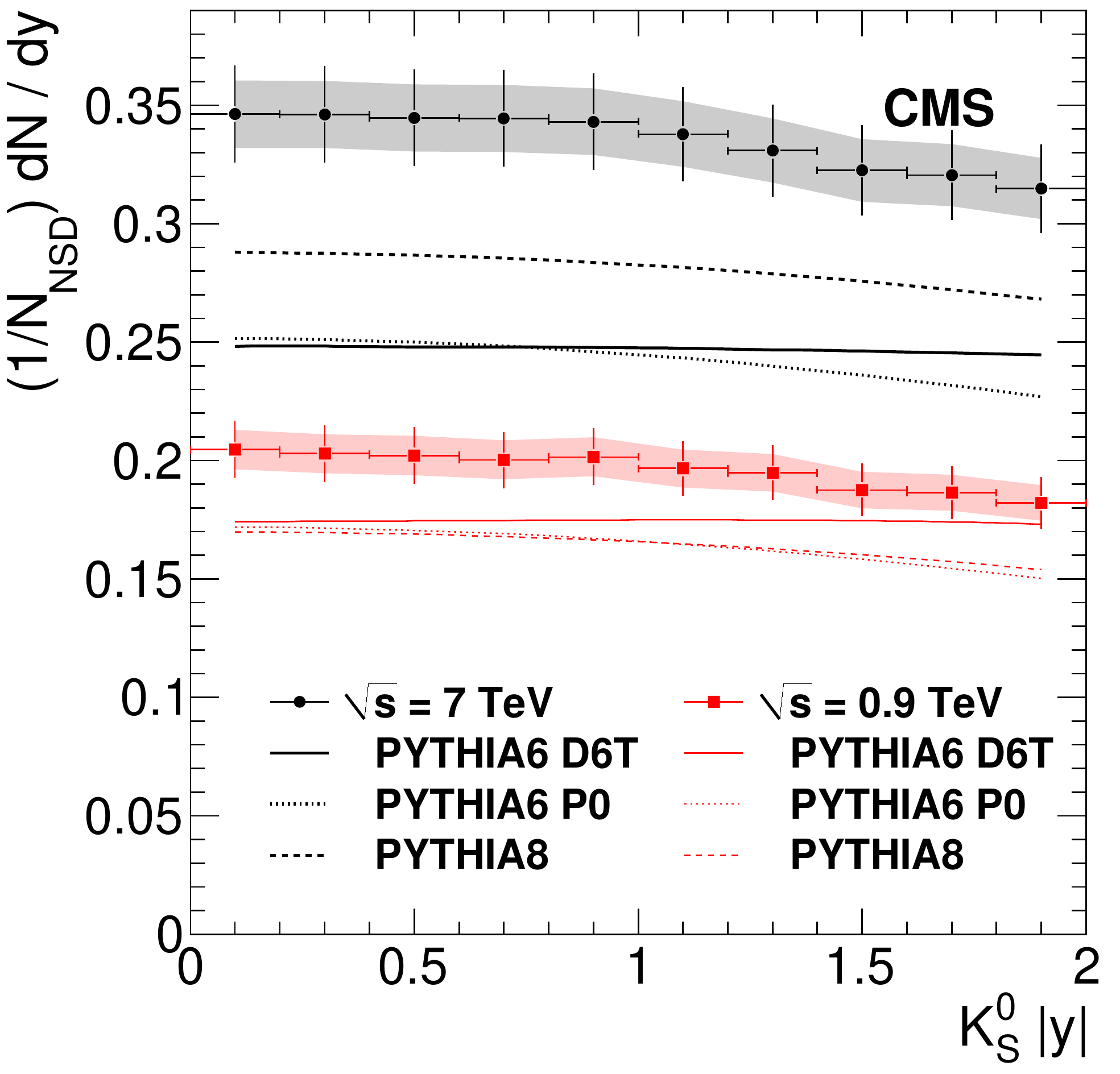}
    \includegraphics[width=0.42\textwidth]{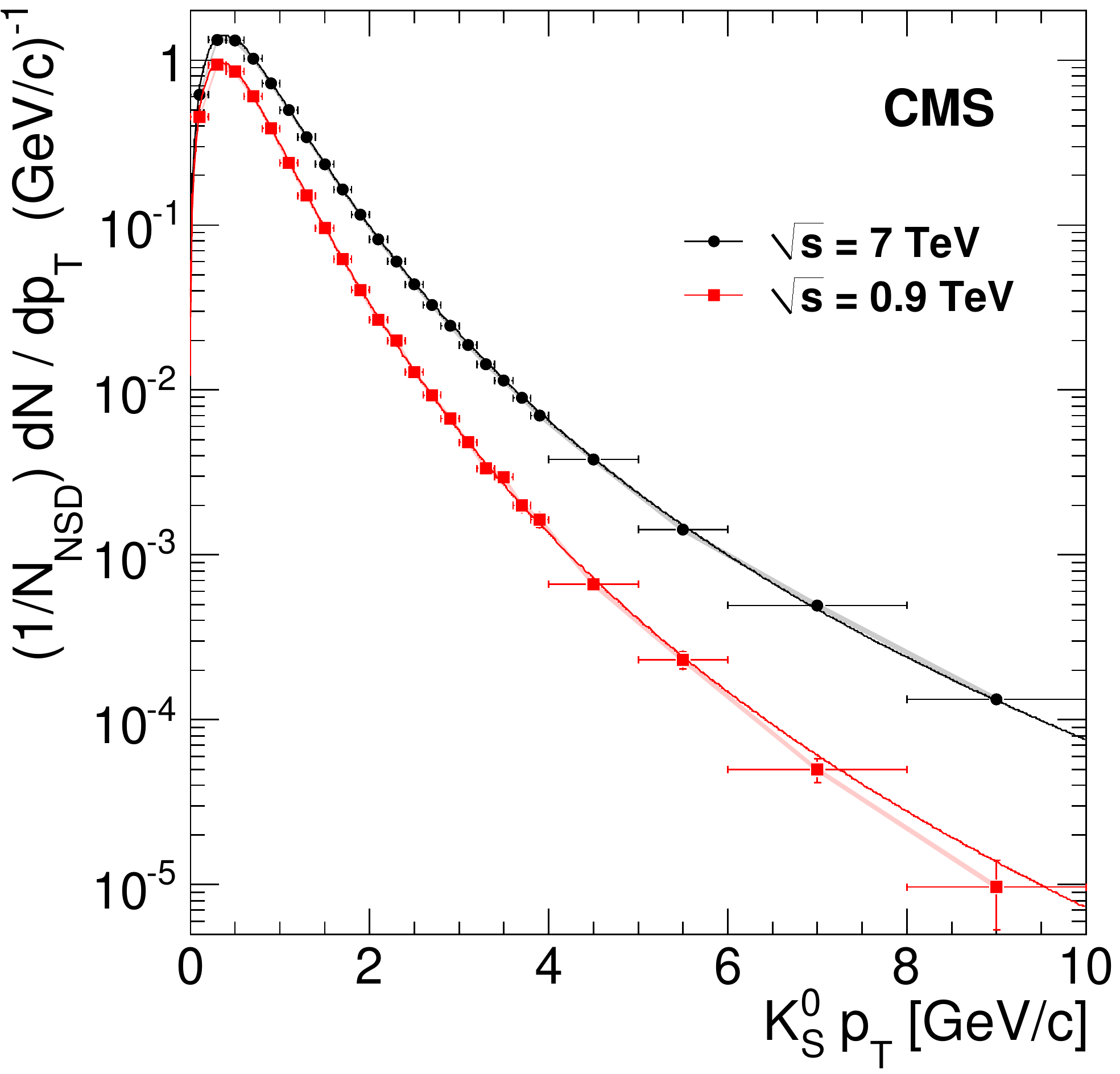}
    \includegraphics[width=0.42\textwidth]{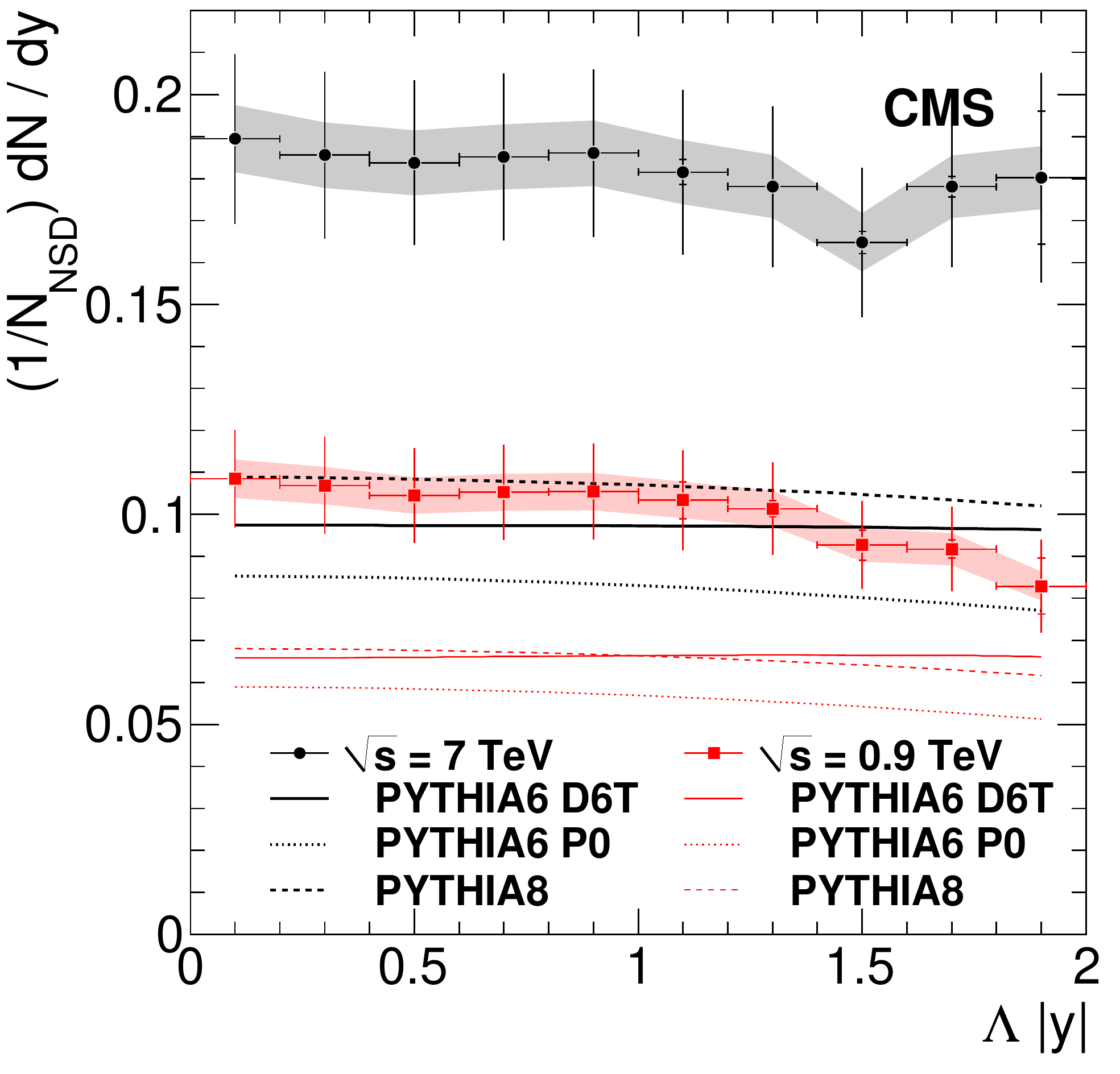}
    \includegraphics[width=0.42\textwidth]{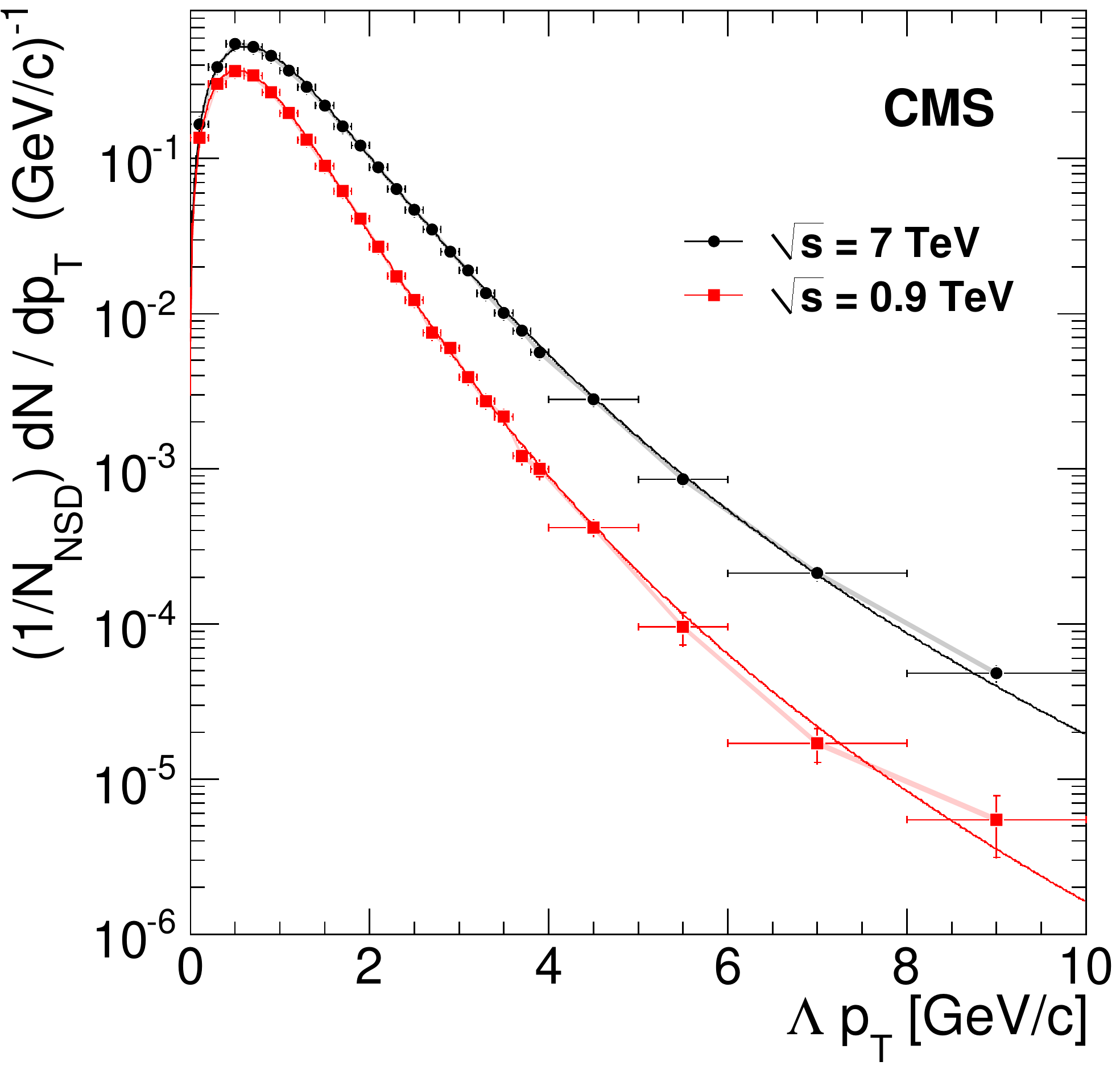}
    \includegraphics[width=0.42\textwidth]{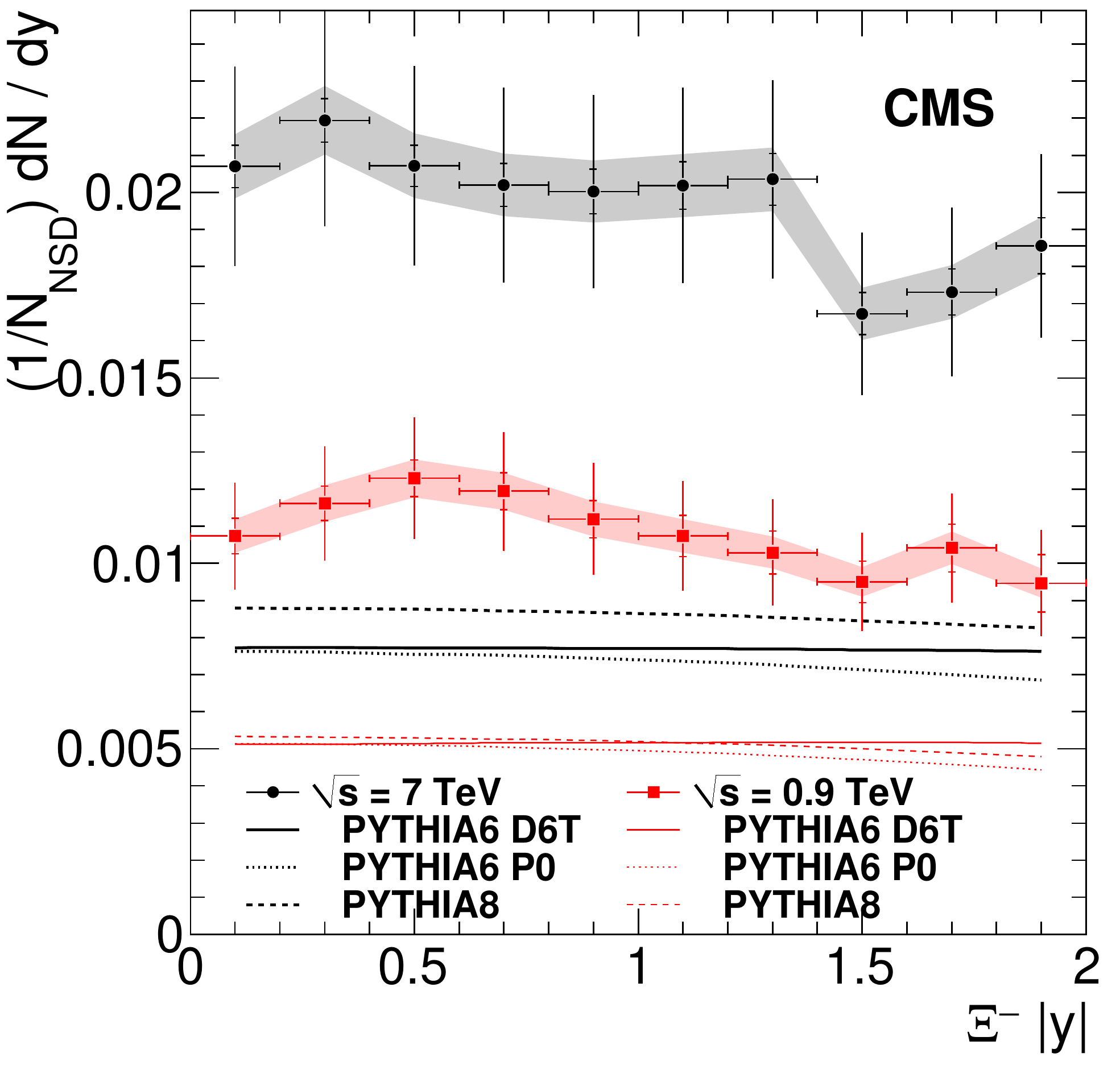}
    \includegraphics[width=0.42\textwidth]{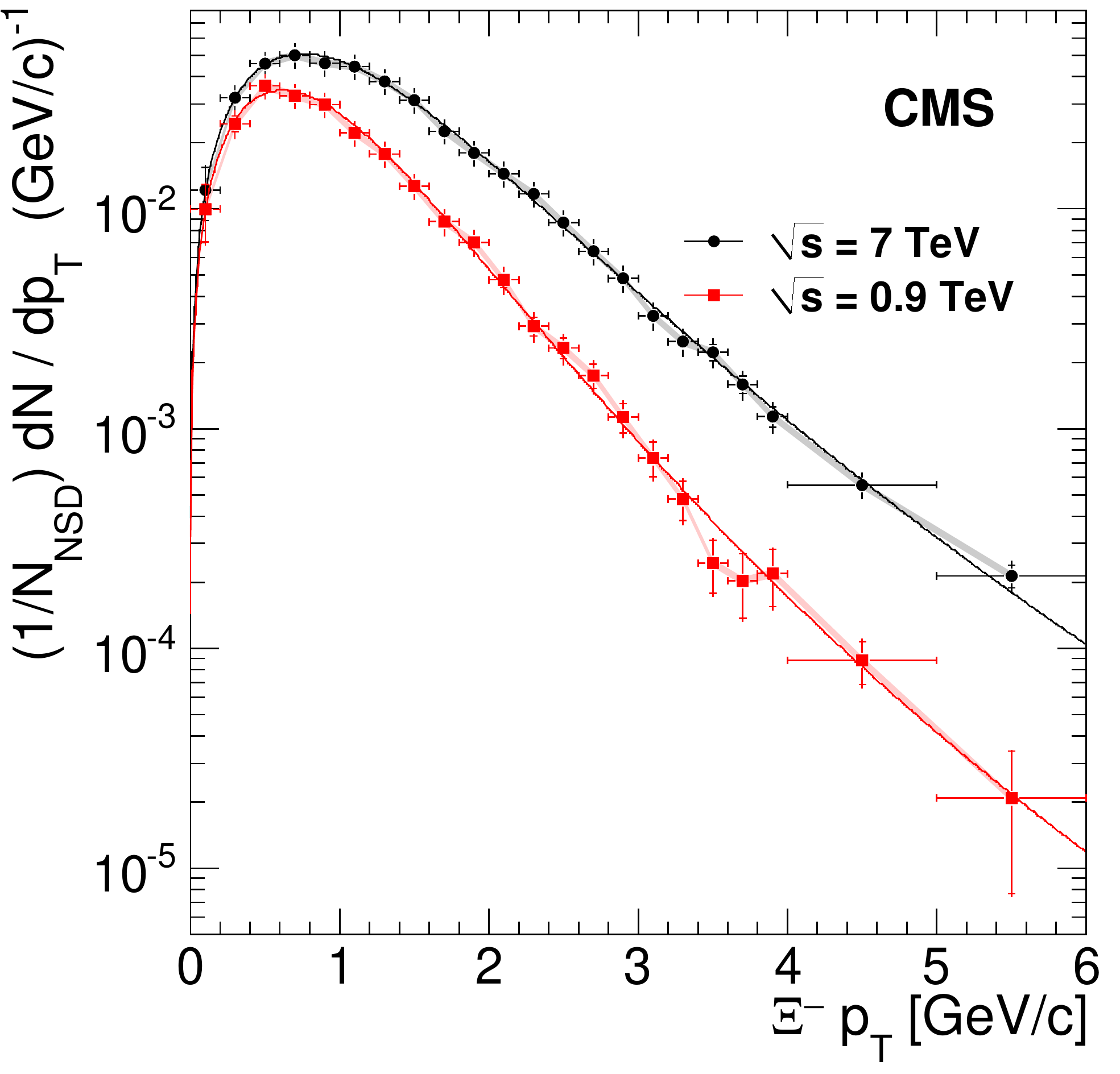}
    \caption{$\mathrm{K}_\mathrm{S}^0$ (top), $\Lambda$ (middle), and $\Xi^-$ (bottom) production
per NSD event versus $|y|$ (left) and \pt (right).
The inner vertical error bars (when visible) show the statistical uncertainties, the outer
the statistical and point-to-point systematic uncertainties summed in quadrature. The
normalization uncertainty is shown as a band. Three \PYTHIA predictions are
overlaid on the $|y|$ distributions. The solid curves in the \pt distributions are fits to
the Tsallis function as described in the text.}
    \label{fig:yptdist}
  \end{center}
\end{figure}

The corrected yields of $\mathrm{K}_\mathrm{S}^0$, $\Lambda$, and $\Xi^-$,
versus $|y|$ and \pt are plotted in Fig.~\ref{fig:yptdist}, normalized to the
number of NSD events.
The rapidity distribution is flat at central rapidities with a slight decrease at higher
rapidities while the \pt distribution is observed to be rapidly falling.
The rapidity distributions also show results from three different \PYTHIA models:
\PYTHIA6.422 with the D6T and P0 tunes\,\cite{Bartalini:2010su,Skands:2009zm} and \PYTHIA8.135\,\cite{Pythia8}.
Fits to the Tsallis function, described below, are overlaid on the \pt distributions.

\subsection{Analysis of \pt spectra}
The corrected $\pt$ spectra are fit to the Tsallis function\,\cite{Tsallis}, as was done for
charged particles\,\cite{CMS_QCD-09-010,CMS_QCD-10-006}.  The Tsallis function used is:
\begin{equation}
\frac{1}{N_{\mathrm{NSD}}} \,\frac{dN}{d\pt} = C\frac{(n-1)(n-2)}{nT[nT+m(n-2)]} \pt\left[1+ \frac{\sqrt{p_\mathrm{T}^2+m^2}-m}{nT}\right]^{-n},
\end{equation}
where $C$ is a normalization parameter and $T$ and $n$ are the shape parameters.  The
results of the fits are shown in Table~\ref{tab:tsallis}.  The data points used in the
fits include only the statistical uncertainty.  The statistical uncertainties on the fit
parameters are obtained from the fit.  The systematic uncertainties are obtained by
varying the cuts and Monte Carlo conditions (tune, material, beamspot, and alignment) in
the same way as used to obtain the point-to-point systematic uncertainties on the
distributions.  The systematic uncertainty on the normalization parameter $C$ also includes
the normalization uncertainty given in Table~\ref{tab:syst}. 
The normalized $\chi^2$ indicates good fits to most of the samples.
The $T$ parameter can be associated
with the inverse slope parameter of an exponential which dominates at low $\pt$, while the
$n$ parameter controls the power law behaviour at high $\pt$.  While both parameters are
necessary, they are highly correlated, with correlation coefficients around 0.9, making it
difficult to elucidate information.  Nevertheless, it is clear that $T$ increases with
particle mass and centre-of-mass energy.  This indicates a broader low-$\pt$ shape at
higher centre-of-mass energy and for higher mass particles.  In contrast, the high $\pt$
power-law behaviour seems to show a much steeper fall off for the two baryons than for the
$\mathrm{K}_\mathrm{S}^0$.  While the power-law behaviour of the baryons does not show any
dependence on the centre-of-mass energy, the fall off of the $\mathrm{K}_\mathrm{S}^0$
particles produced at $\sqrt{s} =0.9$\,TeV is steeper than those produced at
$\sqrt{s}=7$\,TeV\@.

\begin{table}[htbp]
\begin{center}
\caption{\label{tab:tsallis} Results of fitting the Tsallis function to the data.
In the $C$, $T$, and $n$ columns, the first uncertainty is statistical and the second is systematic.
The parameter values and $\chi^2/\mathrm{NDF}$ are obtained from fits to the data with only the
statistical uncertainty included.}
\begin{tabular}{cccccc}\hline
Particle & $\sqrt{s}$ (TeV) & $C$ & $T$ (MeV) & $n$ & $\chi^2/\mathrm{NDF}$ \\\hline
$\mathrm{K}_\mathrm{S}^0$ & 0.9 & $0.776 \pm 0.002 \pm 0.042$ & $187 \pm 1 \pm 4$  & $7.79 \pm 0.07 \pm 0.26$ & $19/21$	\\
$\Lambda$                 & 0.9 & $0.395 \pm 0.002 \pm 0.041$ & $216 \pm 2 \pm 11$ & $9.3 \pm 0.2  \pm 1.1$   & $32/21$	\\
$\Xi^-$                   & 0.9 & $0.043 \pm 0.001 \pm 0.006$ & $250 \pm 8 \pm 48$ & $10.1 \pm 0.9 \pm 4.7$   & $19/19$ \\
$\mathrm{K}_\mathrm{S}^0$ & 7   & $1.329 \pm 0.001 \pm 0.062$ & $220 \pm 1 \pm 3$  & $6.87 \pm 0.02 \pm 0.09$ & $50/21$	\\
$\Lambda$                 & 7   & $0.696 \pm 0.002 \pm 0.058$ & $292 \pm 1 \pm 10$ & $9.3 \pm 0.1  \pm 0.5$   & $128/21$ \\
$\Xi^-$                   & 7   & $0.080 \pm 0.001 \pm 0.012$ & $361 \pm 7 \pm 72$ & $11.2 \pm 0.7 \pm 4.9$   & $21/19$ \\
\end{tabular}
\end{center}
\end{table}

\begin{figure}[hbtp]
  \begin{center} \includegraphics[width=0.6\textwidth]{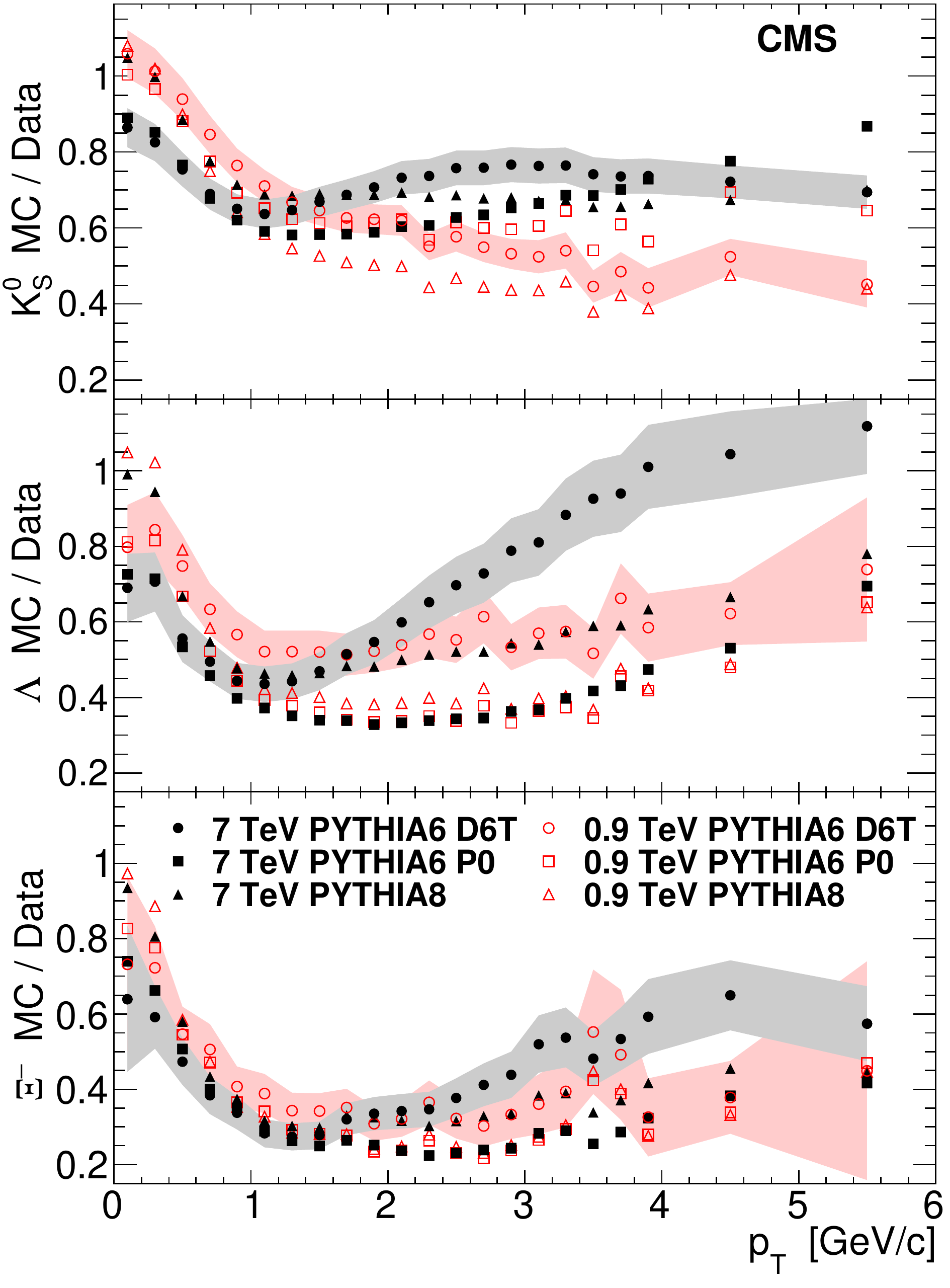}
    \caption{Ratio of MC production to data production of $\mathrm{K}_\mathrm{S}^0$ (top),
    $\Lambda$ (middle), and $\Xi^-$ (bottom) versus $\pt$ at $\sqrt{s} = 0.9$\,TeV (open
    symbols) and $\sqrt{s} = 7$\,TeV (filled symbols).  Results are shown for three
    \PYTHIA predictions at each centre-of-mass energy.  To reduce clutter, the
    uncertainty, shown as a band, is included for only one of the predictions (D6T) at
    each energy.  This uncertainty includes the statistical and point-to-point systematic
    uncertainties added in quadrature but does not include the normalization systematic
    uncertainty.}  \label{fig:ptmc} \end{center}
\end{figure}

We calculate the average $\pt$ directly from the data in the $dN/d\pt$ histograms.  The
Tsallis function fit is used to obtain the correct bin centre and to account for events
beyond the measured $\pt$ range, both of which are small effects.  The statistical
uncertainty on the average \pt is obtained by finding the standard deviation of \pt and dividing by the square root
of the equivalent number of background-free events, where the equivalent number of background-free
events is given by the square of the inverse of the relative uncertainty on the total
number of signal events.  The systematic uncertainty is composed of two components added
in quadrature.  The first component is the same as used in determining the Tsallis
function systematic uncertainties (varying the cuts and Monte Carlo conditions).  The
second component is obtained by using the mean $\pt$ of the fitted Tsallis function.  The
average \pt from data and \PYTHIA6 with the D6T underlying event tune is shown in
Table~\ref{tab:meanpt}.  The \PYTHIA values are quite close to the $\sqrt{s} = $ 7\TeV
data and somewhat lower than the $\sqrt{s} =$ 0.9\TeV data.  Although the average \pt
results from \PYTHIA are relatively close to the data, the \PYTHIA \pt distributions are
significantly broader than the data distributions.  This disagreement can be seen in
Fig.~\ref{fig:ptmc}, which shows the ratio of \PYTHIA to data for production of
$\mathrm{K}_\mathrm{S}^0$, $\Lambda$, and $\Xi^-$ versus transverse momentum.  As well as
a broader distribution, the \PYTHIA distributions also show significant variation as a
function of tune and version.

\begin{table}[htbp]
\begin{center}
\caption{\label{tab:meanpt} Average $\pt$ in units of \MeVc obtained from the appropriate
$dN/d\pt$ distribution as described in the text.  Results from \PYTHIA6 with tune D6T are also
given.  In each data column, the first uncertainty is statistical and the second is systematic.}
\begin{tabular}{c|cc|cc}\hline
         & \multicolumn{2}{c|}{$\sqrt{s} = 0.9\TeV$} & \multicolumn{2}{c}{$\sqrt{s} = 7\TeV$} \\
Particle & Data & MC (D6T) &  Data & MC (D6T) \\  \hline
$\mathrm{K}_\mathrm{S}^0$ & 654 $\!\pm\!$ 1 $\!\pm\!$ 8 & 580 & 790 $\!\pm\!$ 1 $\!\pm\!$ 9 & 757 \\
$\Lambda$                 & 837 $\!\pm\!$ 6 $\!\pm\!$ 40 & 750 & 1037 $\!\pm\!$ 5 $\!\pm\!$ 63 & 1071 \\
$\Xi-$                    & 971 $\!\pm\!$ 14 $\!\pm\!$ 43 & 831 & 1236 $\!\pm\!$ 11 $\!\pm\!$ 72 & 1243 \\
\end{tabular}
\end{center}
\end{table}

\begin{figure}[hbtp]
  \begin{center} \includegraphics[width=0.47\textwidth]{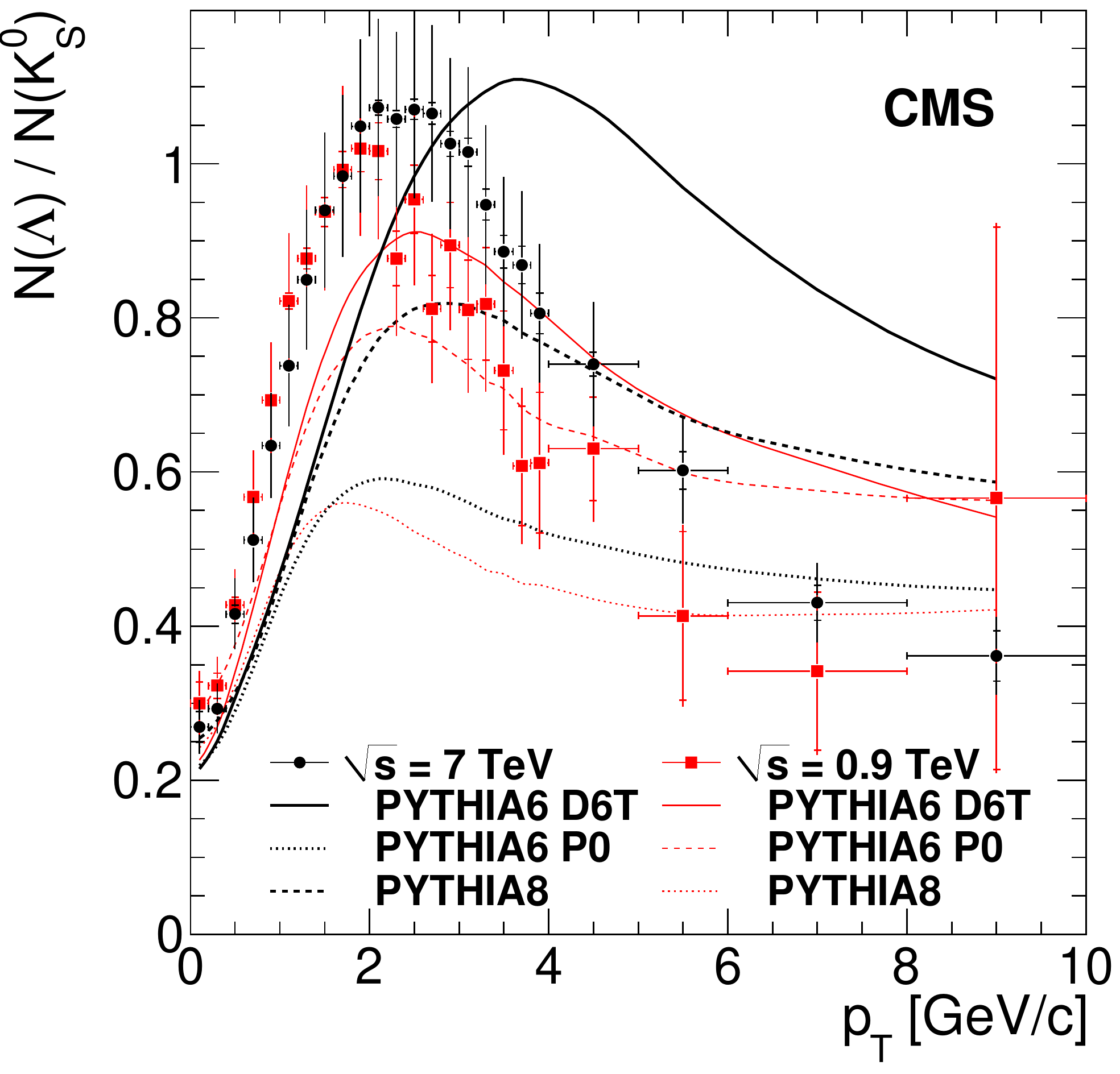}
    \includegraphics[width=0.47\textwidth]{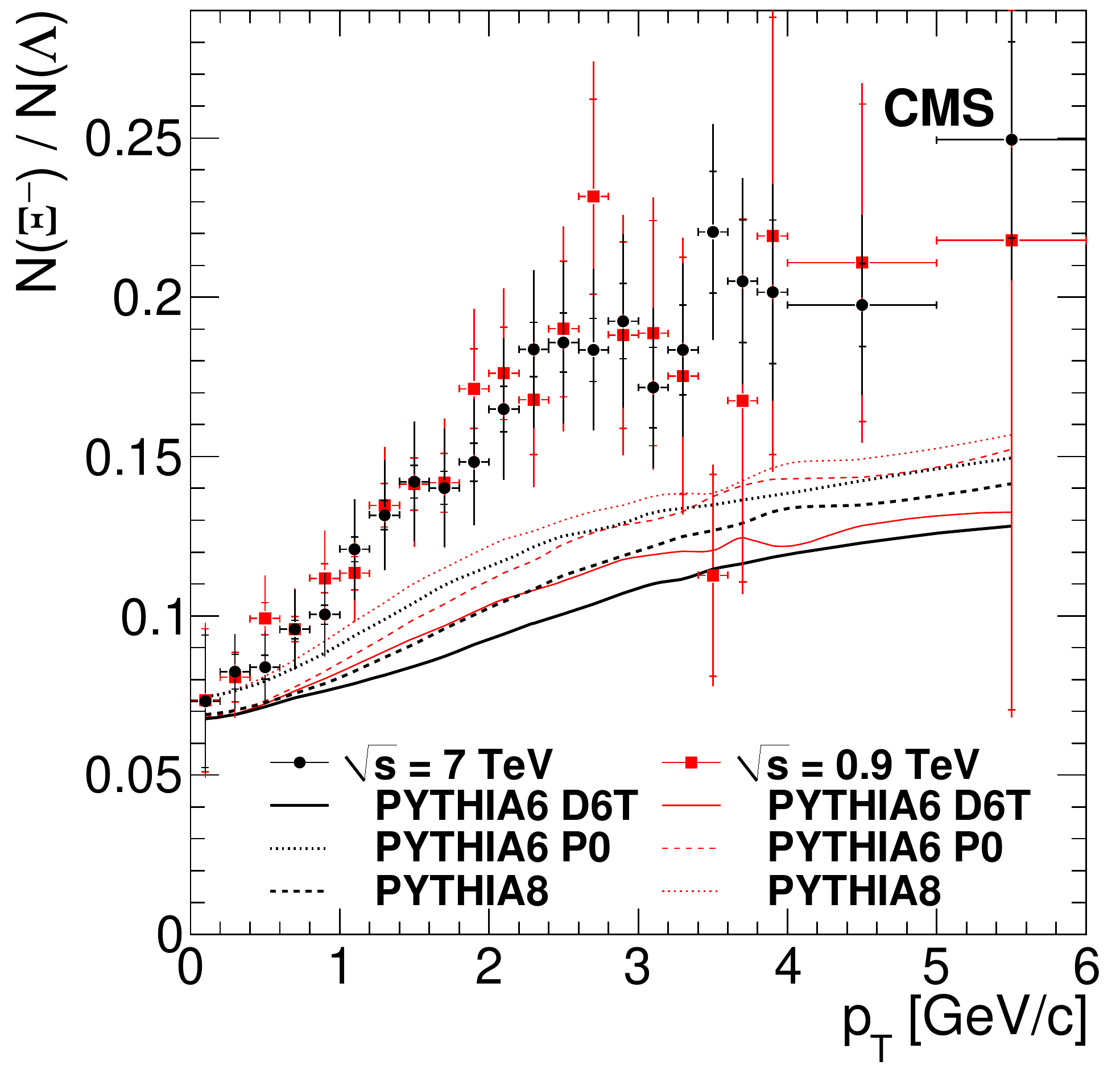}
    \caption{$N(\Lambda)/N(\mathrm{K}_\mathrm{S}^0)$ (left) and $N(\Xi^-)/N(\Lambda)$
    (right) in NSD events versus $\pt$.  The inner vertical error bars (when visible) show
    the statistical uncertainties, the outer the statistical and all systematic
    uncertainties summed in quadrature.  Results are shown for three \PYTHIA predictions
    at each centre-of-mass energy.}  \label{fig:ptratios} \end{center}
\end{figure}

The relative production versus transverse momentum between different species is shown in
Fig.~\ref{fig:ptratios}.  The $N(\Lambda)/N(\mathrm{K}_\mathrm{S}^0)$ and
$N(\Xi^-)/N(\Lambda)$ distributions both increase with $\pt$ at low $\pt$, as expected
from the higher average $\pt$ for the higher mass particles.  At higher $\pt$ the
$N(\Lambda)/N(\mathrm{K}_\mathrm{S}^0)$ distribution drops off while the
$N(\Xi^-)/N(\Lambda)$ distribution appears to plateau.  This is consistent with the values
of the power-law parameter $n$ for these distributions.  Interestingly, the collision
energy has no observable effect on the level or shape of these production ratios.  The
\PYTHIA results are superimposed on the same plot.  While \PYTHIA reproduces the general
features, it differs significantly in the details and shows large variations depending on
tune and version.

\begin{figure}[hbtp]
  \begin{center} \includegraphics[width=0.7\textwidth]{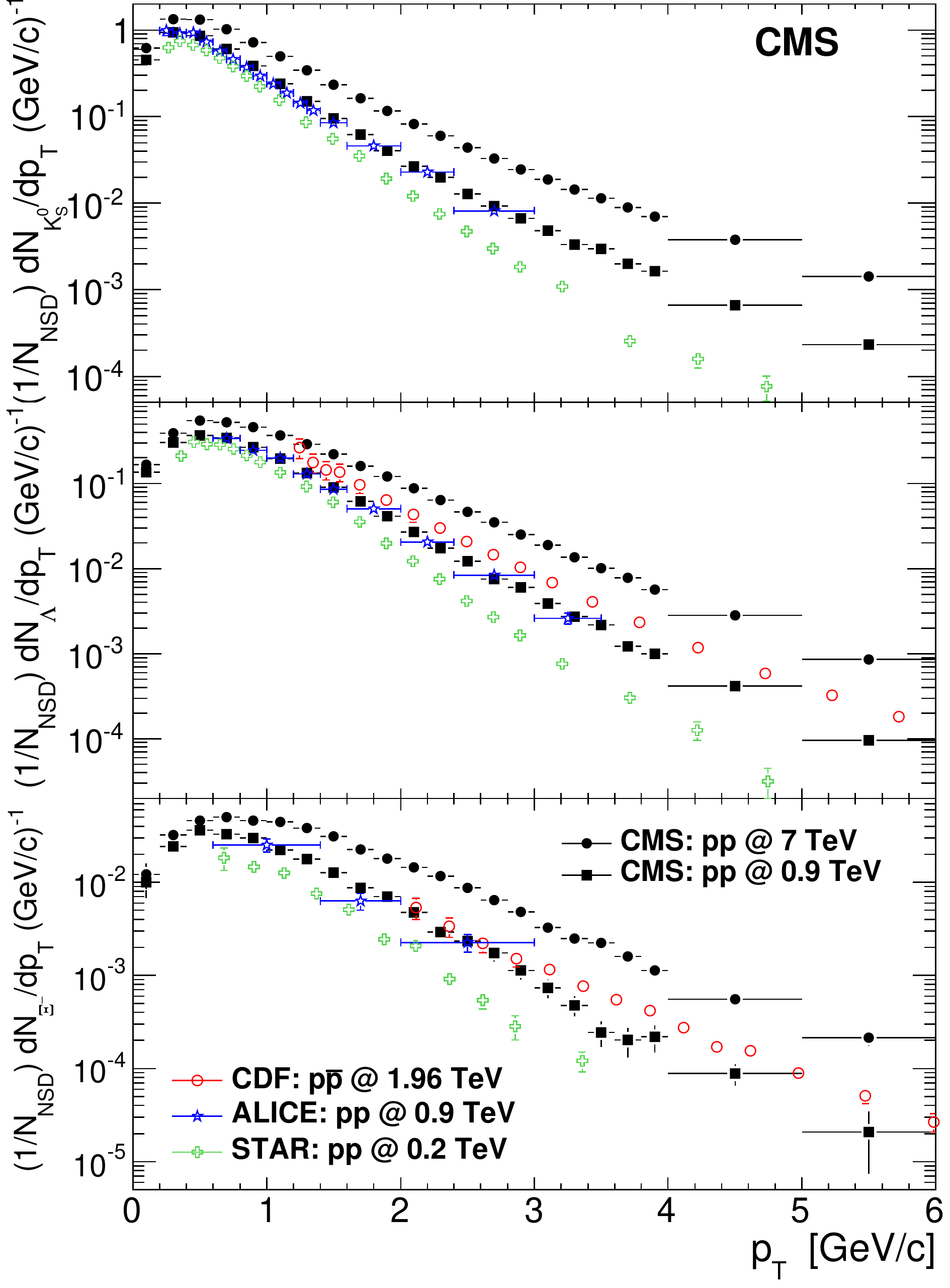}
    \caption{$\mathrm{K}_\mathrm{S}^0$ (top), $\Lambda$ (middle), and $\Xi^-$ (bottom)
    production per event versus \pt.  The error bars on the CMS results show the combined
    statistical, point-to-point systematic, and normalization systematic uncertainties.  The
    error bars on the CDF\,\cite{Aaltonen:2011wz}, ALICE\,\cite{AliceStrange}, and
    STAR\,\cite{Abelev:2006cs} results show the combined statistical and systematic
    uncertainties.  The CMS, CDF, and STAR results are normalized to NSD events while the
    ALICE results are normalized to all inelastic events.}  \label{fig:ptdata}
    \end{center}
\end{figure}

\begin{figure}[hbtp]
  \begin{center} \includegraphics[width=0.7\textwidth]{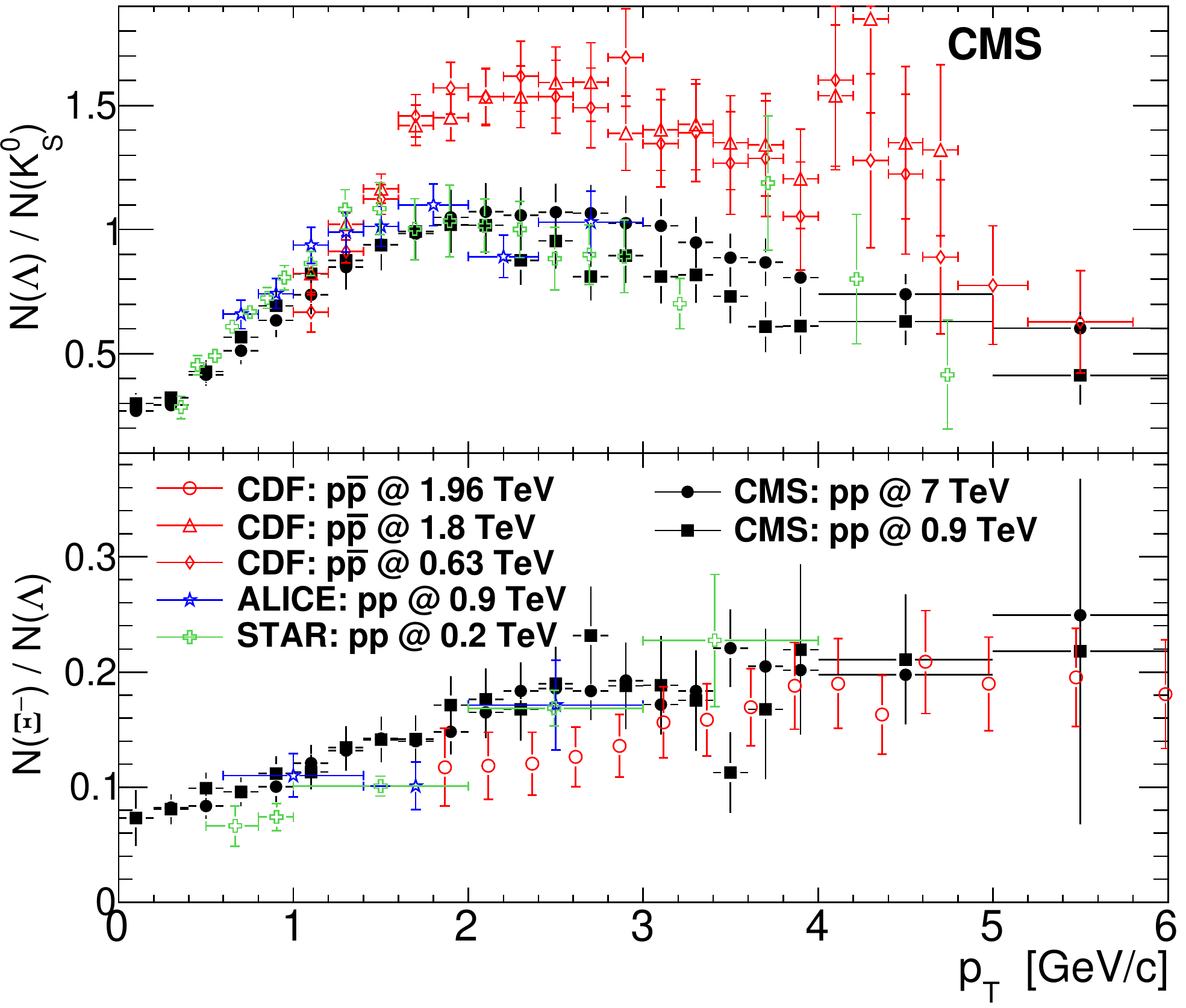}
    \caption{Ratio of $\Lambda$ to $\mathrm{K}_\mathrm{S}^0$ production (top) and
    $\Xi^-$ to $\Lambda$ production (bottom) versus $\pt$.  The CMS,
    ALICE\,\cite{AliceStrange}, and STAR\,\cite{Abelev:2006cs} error bars include the
    statistical and systematic uncertainties.  The CDF error bars include the statistical
    uncertainties for $N(\Lambda)/N(\mathrm{K}_\mathrm{S}^0)$\,\cite{Acosta:2005pk} and
    the statistical and systematic uncertainties for
    $N(\Xi^-)/N(\Lambda)$\,\cite{Aaltonen:2011wz}.  The CDF
    $N(\Lambda)/N(\mathrm{K}_\mathrm{S}^0)$ bin sizes are doubled to reduced fluctuations.
    For experiments in which the binning for $\Lambda$ and $\Xi^-$ is different (ALICE and
    STAR), bins are merged to provide common bin ranges in the $N(\Xi^-)/N(\Lambda)$
    distribution.}  \label{fig:ptratdata} \end{center}
\end{figure}

Figure~\ref{fig:ptdata} shows a comparison of the CMS $\pt$ distributions with results
from other recent experiments\,\cite{Abelev:2006cs,AliceStrange,Aaltonen:2011wz}.  To
compare with the CMS results, the CDF, ALICE, and STAR distributions are multiplied by
$8\pi \pt$, $4$, and $8\pi$, respectively.  The CDF cross sections are also divided by
$49$~mb (the NSD cross section used by CDF\,\cite{Aaltonen:2011wz}) to obtain distributions normalized to NSD
events, matching the CMS and STAR normalization.  The ALICE results are normalized to
inelastic events (including single diffractive events).  The ALICE and CMS results at
0.9~TeV agree for all three particles.  The distributions behave as expected, with
higher centre-of-mass energy corresponding to increased production rates and harder
spectra.  To remove the effect of normalization, Fig.~\ref{fig:ptratdata} shows a
comparison of $\Lambda$ to $\mathrm{K}_\mathrm{S}^0$ and $\Xi^-$ to $\Lambda$ production
ratios versus transverse momentum.  The CMS results agree with the results from pp
collisions at $\sqrt{s} = $0.2\,TeV from STAR\,\cite{Abelev:2006cs} and at $\sqrt{s} =
$0.9\,TeV results from ALICE\,\cite{AliceStrange}.  These three results show a remarkable
consistency across a wide variety of collision energies.  In contrast, the CDF values for
$N(\Lambda)/N(\mathrm{K}_\mathrm{S}^0)$\,\cite{Acosta:2005pk} are significantly higher
than the CMS results while the CDF measurements of
$N(\Xi^-)/N(\Lambda)$\,\cite{Aaltonen:2011wz} are lower, albeit with less significance.

\begin{figure}[hbtp]
  \begin{center}
   \includegraphics[width=0.7\textwidth]{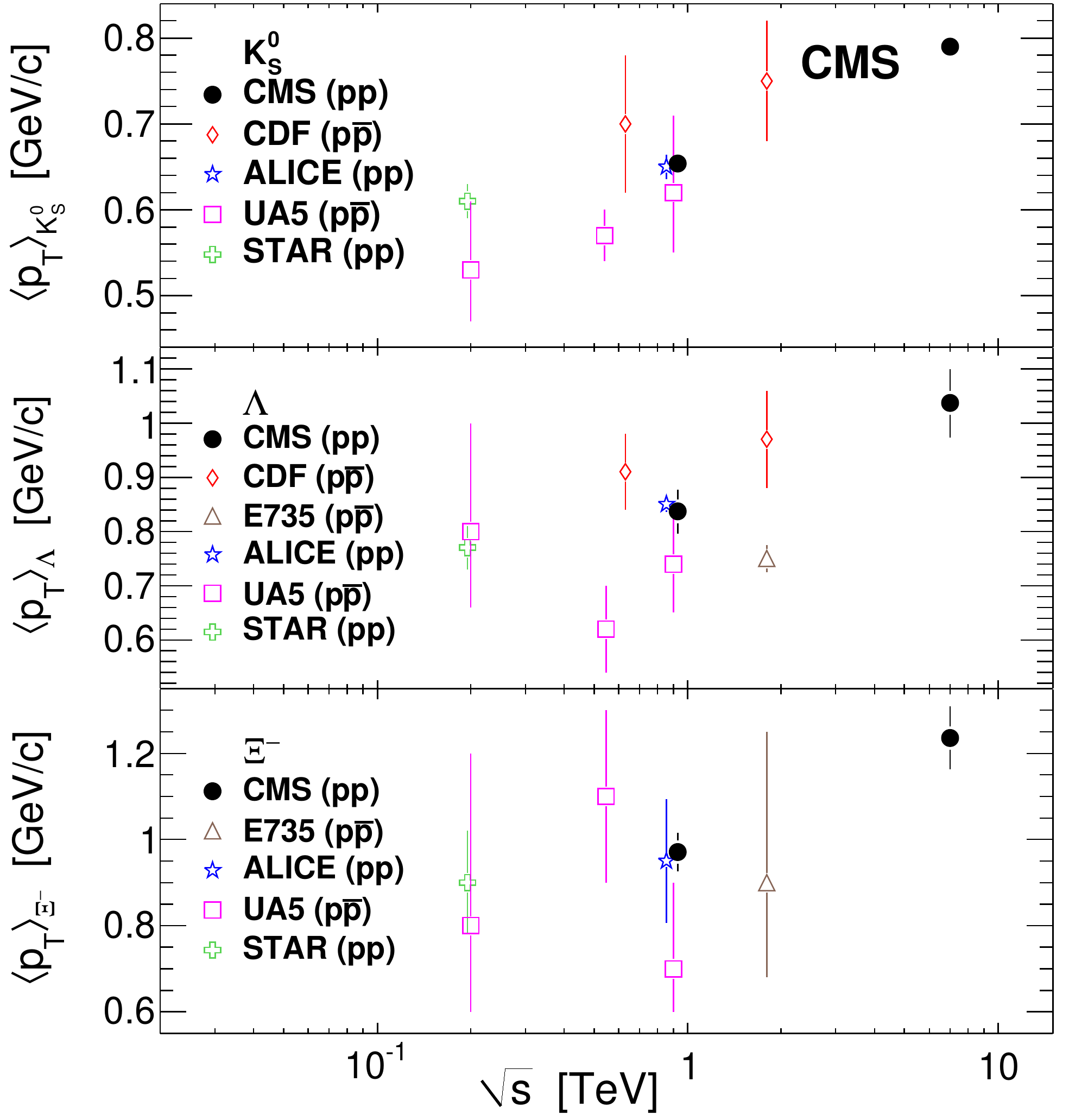}
    \caption{Average \pt for $\mathrm{K}_\mathrm{S}^0$ (top),
$\Lambda$ (middle), and $\Xi^-$ (bottom), as a function of the centre-of-mass energy.
The CMS measurements are for $|y|<2$. The other results are from
UA5\,\cite{Ansorge:1989ba,Alner:1987wb,Ansorge:1988fq,Alner:1985ra,Alner:1984qa} ($\mathrm{p}\bar{\mathrm{p}}$
collisions covering $|y|<2.5$, $|y|<2$, and $|y|<3$ for $\mathrm{K}_\mathrm{S}^0$, $\Lambda$, and $\Xi^-$, respectively),
E735\,\cite{Alexopoulos:1992ut} ($\mathrm{p}\bar{\mathrm{p}}$ collisions using tracks with $-0.36<\eta<1.0$),
CDF\,\cite{Acosta:2005pk} ($\mathrm{p}\bar{\mathrm{p}}$ collisions covering $|\eta|<1.0$),
STAR\,\cite{Abelev:2006cs} (pp collisions covering $|y|<0.5$), and
ALICE\,\cite{AliceStrange} (pp collisions covering $|y|<0.75$ for $\mathrm{K}_\mathrm{S}^0$ and $\Lambda$ and $|y|<0.8$ for $\Xi^-$).
Some points have been slightly offset from the true energy to improve visibility.  The vertical bars
indicate the statistical and systematic uncertainties (when available) summed in quadrature.}
    \label{fig:meanpt}
  \end{center}
\end{figure}

Reducing the $\pt$ distributions to a single value, the average $\pt$, we compare the CMS results
with earlier results at lower energies in
Fig.~\ref{fig:meanpt}\,\cite{Ansorge:1989ba,Alner:1987wb,Ansorge:1988fq,Alner:1985ra,Alner:1984qa,
Alexopoulos:1992ut,Acosta:2005pk,Abelev:2006cs,AliceStrange}.  The CMS results are
in excellent agreement with the recent ALICE measurements at 0.9 TeV\@.
The CMS results continue the overall trend of increasing average \pt with increasing particle
mass and increasing centre-of-mass energy.

\subsection{Analysis of production rate}
As a measure of the overall production rate in NSD events, $\frac{dN}{dy}|_{y\approx 0}$ and the total yield
for $|y|<2$ were extracted and tabulated in Table~\ref{tab:perevt}.  The quantity $\frac{dN}{dy}|_{y\approx 0}$
is the average value of $\frac{dN}{dy}$ over the region $|y|<0.2$.  The integrated yields for $|y|<2$ are
obtained by integrating the $\pt$ spectra, using the Tsallis function fit to account for particles
above the measured $\pt$ range.

\begin{figure}[hbtp]
  \begin{center}
   \includegraphics[width=0.7\textwidth]{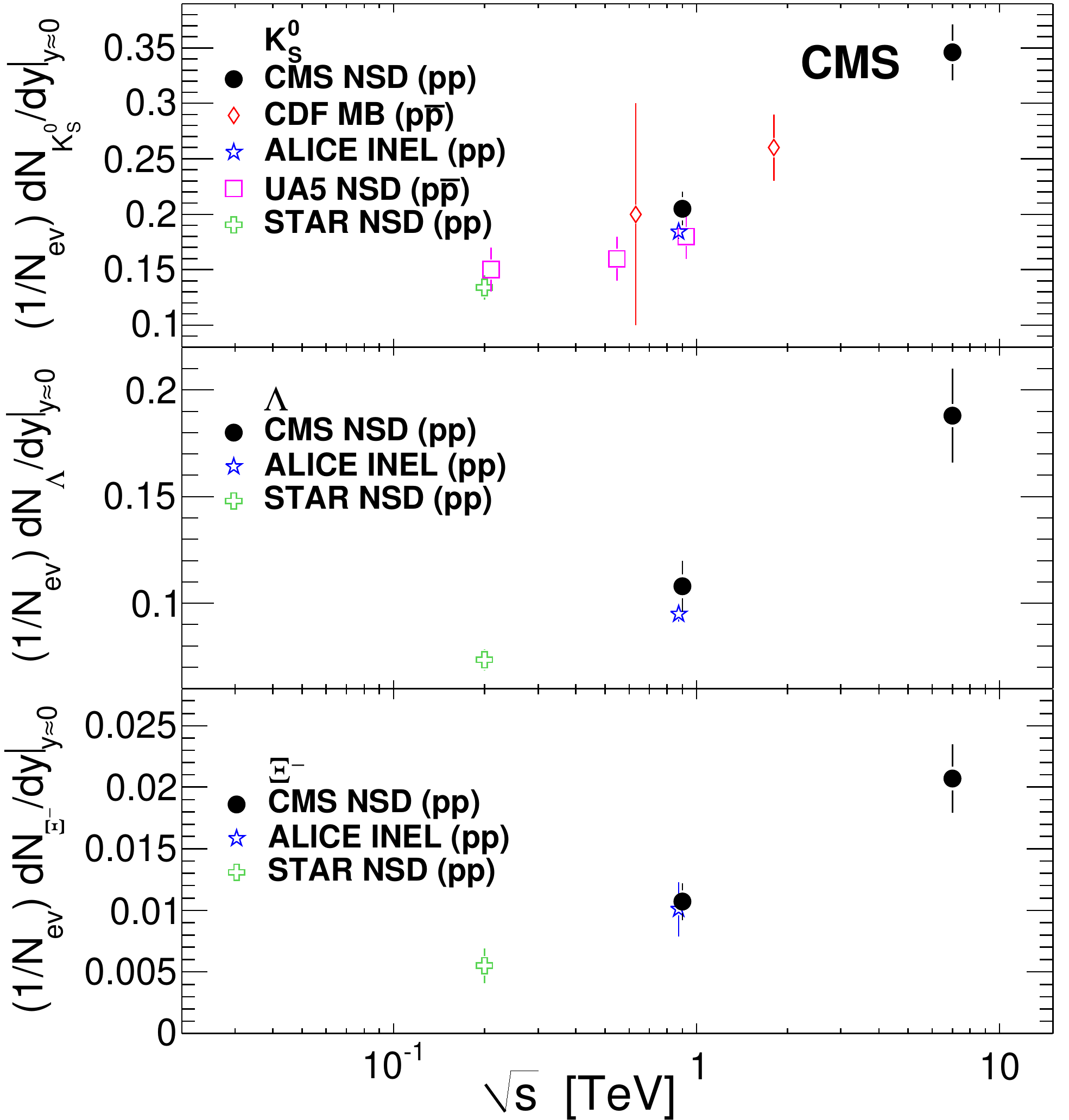}
    \caption{The central rapidity production rate for $\mathrm{K}_\mathrm{S}^0$ (top),
$\Lambda$ (middle), and $\Xi^-$ (bottom), as a function of the centre-of-mass energy.
The previous results are from
UA5\,\cite{Alner:1985ra,Ansorge:1988fq} ($\mathrm{p}\bar{\mathrm{p}}$),
CDF\,\cite{Abe:1989hy} ($\mathrm{p}\bar{\mathrm{p}}$), STAR\,\cite{Abelev:2006cs} (pp),
and ALICE\,\cite{AliceStrange} (pp).  The CMS, UA5, and STAR results are normalized to NSD
events.  The CDF results are normalized to events passing their trigger and event
selection defined chiefly by activity in both sides of the detector, at least four tracks,
and a primary vertex.  The ALICE results are normalized to all inelastic events.  Some
points have been slightly offset from the true energy to improve visibility.  The vertical
bars indicate the statistical uncertainties for the UA5 and CDF results and the combined
statistical and systematic uncertainties for the CMS, ALICE, and STAR results.}
\label{fig:ydatacomp} \end{center}
\end{figure}

The central production rates of $\mathrm{K}_\mathrm{S}^0$, $\Lambda$, and $\Xi^-$ are compared to previous results in
Fig.~\ref{fig:ydatacomp}.  The results show the expected increase in production with centre-of-mass energy
with little evidence of a difference due to beam particles.  As the ALICE results are normalized to all
inelastic collisions, they are expected to be somewhat lower than the CMS results.

\begin{figure}[hbtp]
  \begin{center} \includegraphics[width=0.45\textwidth]{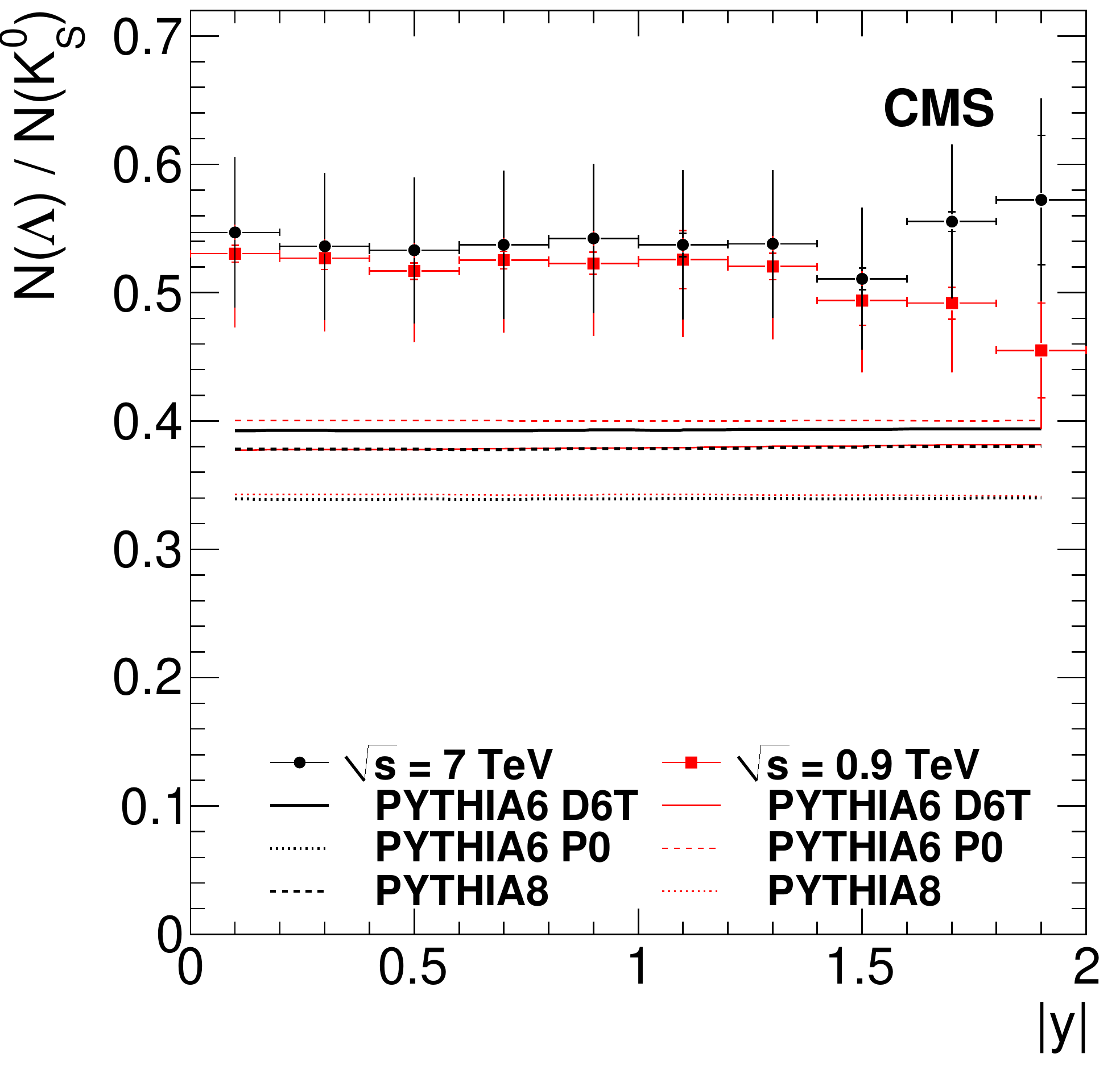}
    \includegraphics[width=0.45\textwidth]{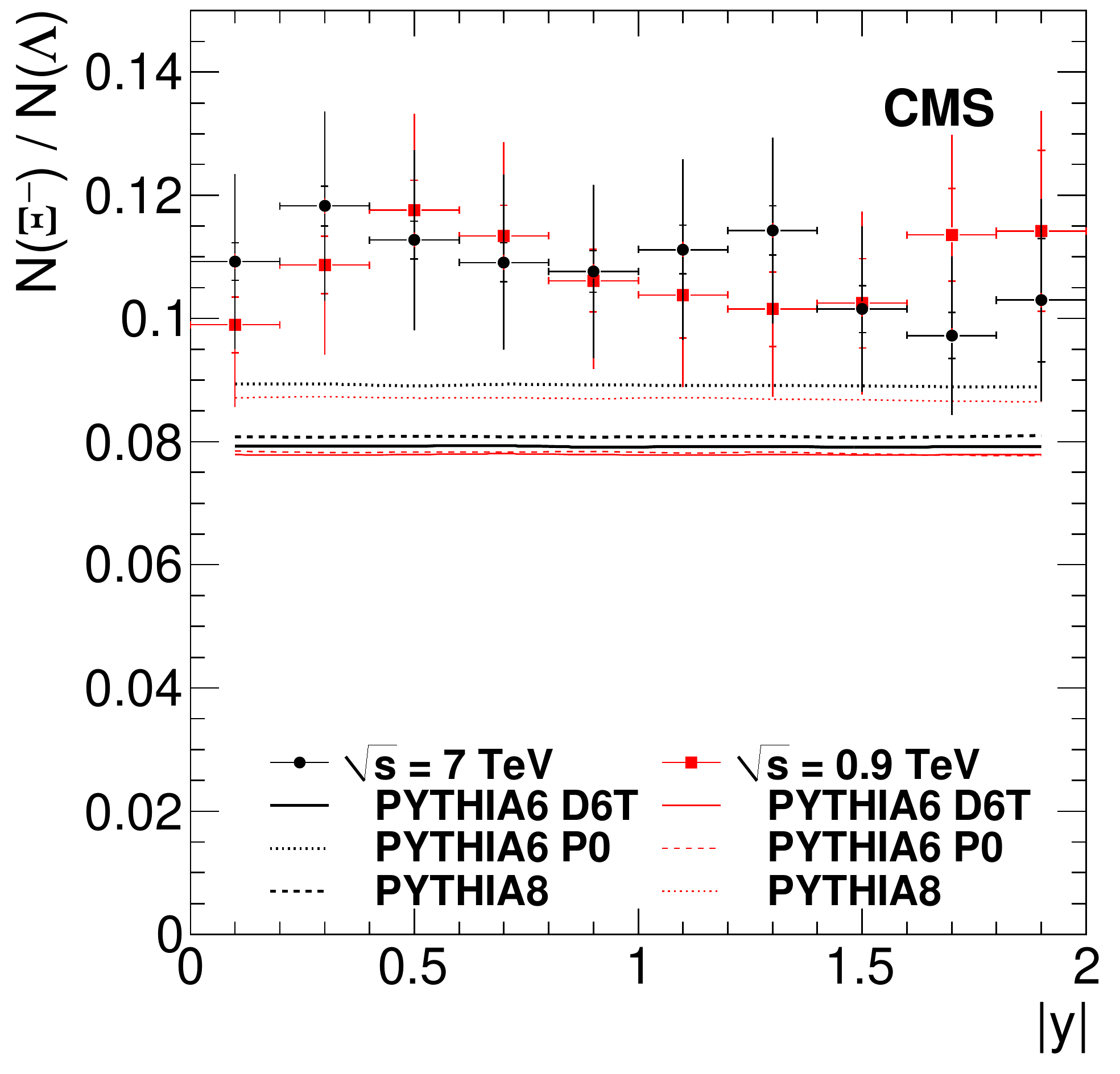} \caption{The
    production ratios $N(\Lambda)/N(\mathrm{K}_\mathrm{S}^0)$ (left) and
    $N(\Xi^-)/N(\Lambda)$ (right) in NSD events versus $|y|$.  The inner vertical error
    bars (when visible) show the statistical uncertainties, the outer the statistical and
    all systematic uncertainties summed in quadrature.  Results are shown for three
    \PYTHIA predictions at each centre-of-mass energy.}
\label{fig:yratios} \end{center}
\end{figure}

The production ratios $N(\mathrm{K}_\mathrm{S}^0)/N(\Lambda)$ and $N(\Xi^-)/N(\Lambda)$ versus $|y|$ are shown in
Fig.~\ref{fig:yratios}.  The rapidity distributions are very flat and, as observed in the $\pt$
distributions of Fig.~\ref{fig:ptratios}, show no dependence on centre-of-mass energy.  Three \PYTHIA
predictions at each centre-of-mass energy are also shown in Fig.~\ref{fig:yratios}.  These results
confirm what can already be seen in the comparisons shown in the left panes of Fig.~\ref{fig:yptdist};
\PYTHIA underestimates the production of strange particles and the discrepancy grows with particle mass.

\begin{table}[htbp]
\begin{center}
\caption{\label{tab:perevt} $\frac{dN}{dy}|_{y\approx 0}$ and integrated yields $(|y|<2.0)$ per NSD event from data.
In each data column, the first uncertainty is statistical and the second is systematic.}
\begin{tabular}{@{}c|c@{~~~}c|c@{~~~}c@{}}\hline
         & \multicolumn{2}{c|}{$\sqrt{s} = 0.9\TeV$} & \multicolumn{2}{c}{$\sqrt{s} = 7\TeV$} \\
Particle$\!$ & $\frac{dN}{dy}|_{y\approx 0}$ & $N$ & $\frac{dN}{dy}|_{y\approx 0}$ & $N$ \\ \hline
$\mathrm{K}_\mathrm{S}^0$ & 0.205 $\!\pm\!$ 0.001 $\!\pm\!$ 0.015 & 0.784 $\!\pm\!$ 0.002 $\!\pm\!$ 0.056 & 0.346 $\!\pm\!$ 0.001 $\!\pm\!$ 0.025 & 1.341 $\!\pm\!$ 0.001 $\!\pm\!$ 0.097 \\
$\Lambda$                 & 0.108 $\!\pm\!$ 0.001 $\!\pm\!$ 0.012 & 0.404 $\!\pm\!$ 0.004 $\!\pm\!$ 0.046 & 0.189 $\!\pm\!$ 0.001 $\!\pm\!$ 0.022 & 0.717 $\!\pm\!$ 0.005 $\!\pm\!$ 0.082 \\
$\Xi^-$                   & 0.011 $\!\pm\!$ 0.001 $\!\pm\!$ 0.001 & 0.043 $\!\pm\!$ 0.001 $\!\pm\!$ 0.006 & 0.021 $\!\pm\!$ 0.001 $\!\pm\!$ 0.003 & 0.080 $\!\pm\!$ 0.001 $\!\pm\!$ 0.011 \\
\end{tabular}
\end{center}
\end{table}

Table~\ref{tab:mccomp} shows a comparison of the production rate of data to \PYTHIA6 with
the D6T tune.  The left column shows a large increase in the strange particle production
cross section as the centre-of-mass energy increases from 0.9 to 7~TeV\@.  The systematic
uncertainties for this ratio are reduced as the same uncertainty affects both samples
nearly equally.  The results for $\mathrm{K}_\mathrm{S}^0$ and $\Lambda$ are consistent
with the increase observed in inclusive charged particle
production\,\cite{CMS_QCD-09-010,CMS_QCD-10-006} $(\frac{5.82}{3.48} = 1.67)$
while the $\Xi^-$ results show a slightly greater increase.  The increase in particle
production from 0.9 to 7~TeV is not well modelled by \PYTHIA6.  Another feature, seen in
the right column, is the deficit of strange particles produced by \PYTHIA6.  The deficit
of $\mathrm{K}_\mathrm{S}^0$ particles in the MC, 15\% (28\%) low at 0.9 (7)~TeV, is
consistent with the results found in the production of charged
particles\,\cite{CMS_QCD-09-010,CMS_QCD-10-006}.  However, the deficit is much worse as
the mass increases, resulting in a 63\% reduction in $\Xi^-$ particles in MC compared to
data at $\sqrt{s} = $7\TeV.  While values are only presented for \PYTHIA6 with the D6T
tune, the same features are also evident for the other two \PYTHIA comparisons in
the rapidity distribution plots in Fig.~\ref{fig:yptdist}.

\begin{table}[htbp]
\begin{center}
\caption{\label{tab:mccomp} Comparison of strangeness production rates between \PYTHIA6 Monte Carlo (D6T) and data.  In
each column, the first uncertainty is statistical and the second is systematic.}
\begin{tabular}{c|cc|cc}
Particle    & \multicolumn{2}{c|}{$\displaystyle \left[ \frac{\frac{dN}{dy}|_{y\approx 0}(7\TeV)}{\frac{dN}{dy}|_{y\approx 0}(0.9\TeV)} \right]$} &
\multicolumn{2}{c}{$\displaystyle \left[ \frac{\frac{dN}{dy}|_{y\approx 0}(\mathrm{MCD6T})}{\frac{dN}{dy}|_{y\approx 0}(\mathrm{Data})} \right]$} \\[3pt]
& Data & MC (D6T) & $\sqrt{s} = $ 0.9\TeV & $\sqrt{s} = $ 7\TeV  \\ \hline
$\mathrm{K}_\mathrm{S}^0$ & 1.69 $\pm$ 0.01 $\pm$ 0.06 & 1.42 & 0.852 $\pm$ 0.005 $\pm$ 0.061 & 0.717 $\pm$ 0.001 $\pm$ 0.052 \\
$\Lambda$                 & 1.75 $\pm$ 0.02 $\pm$ 0.08 & 1.48 & 0.606 $\pm$ 0.007 $\pm$ 0.070 & 0.514 $\pm$ 0.003 $\pm$ 0.059 \\
$\Xi^-$                   & 1.93 $\pm$ 0.10 $\pm$ 0.09 & 1.51 & 0.477 $\pm$ 0.021 $\pm$ 0.064 & 0.373 $\pm$ 0.010 $\pm$ 0.050 \\
\end{tabular}
\end{center}
\end{table}

\section{Conclusions}

This article presents a study of the production of $\mathrm{K}_\mathrm{S}^0$, $\Lambda$,
and $\Xi^-$ particles in proton-proton collisions at centre-of-mass energies 0.9 and 7
TeV\@.  By fully exploiting the low-momentum track reconstruction capabilities of CMS, we
have measured the transverse-momentum distribution of these strange particles down to
zero.  From this sample of 10 million strange particles, the transverse
momentum distributions were measured out to 10~GeV/$c$ for $\mathrm{K}_\mathrm{S}^0$ and
$\Lambda$ and out to 6~GeV/$c$ for $\Xi^-$.  We fit these distributions with a Tsallis
function to obtain information on the exponential decay at low $\pt$ and the power-law
behaviour at high $\pt$.  All species show a flattening of the exponential decay as the
centre-of-mass energy increases.  While the baryons show little change in the high-$\pt$
region, the $\mathrm{K}_\mathrm{S}^0$ power-law parameter decreases from 7.8 to 6.9.  The
average $\pt$ values, calculated directly from the data, are found to increase with
particle mass and centre-of-mass energy, in agreement with predictions and other
experimental results.  While the \PYTHIA $\pt$ distributions used in this analysis show
significant variation based on tune and version, they are all broader than the data
distributions.

We have also measured the production versus rapidity and extracted the value of $dN/dy$ in
the central rapidity region.  The increase in production of strange particles as the
centre-of-mass energy increases from 0.9 to 7~TeV is approximately consistent with the
results for inclusive charged particles.  However, as in the inclusive charged particle
case, \PYTHIA fails to match this increase.  For $\mathrm{K}_\mathrm{S}^0$ production, the
discrepancy is similar to what has been found in charged particles.  However, the deficit
between \PYTHIA and data is significantly larger for the two hyperons at both energies,
reaching a factor of three discrepancy for $\Xi^-$ production at $\sqrt{s} =$ 7\TeV.  If a
quark-gluon plasma or other collective effects were present, we might expect an enhancement
of double-strange baryons to single-strange baryons and/or an enhancement of strange
baryons to strange mesons.  However, the production ratios
$N(\Lambda)/N(\mathrm{K}_\mathrm{S}^0)$ and $N(\Xi^-)/N(\Lambda)$ versus rapidity and
transverse momentum show no change with centre-of-mass energy.  Thus, the deficiency in
\PYTHIA is likely originating from parameters regulating the frequency of strange quarks
appearing in colour strings.  The variety of measurements presented here can be used to
tune \PYTHIA and other models as well as a baseline to understand measurements of
strangeness production in heavy-ion collisions.

\section*{Acknowledgements}
We wish to congratulate our colleagues in the CERN accelerator departments for the
excellent performance of the LHC machine. We thank the technical and administrative staff
at CERN and other CMS institutes, and acknowledge support from: FMSR (Austria); FNRS and
FWO (Belgium); CNPq, CAPES, FAPERJ, and FAPESP (Brazil); MES (Bulgaria); CERN; CAS, MoST,
and NSFC (China); COLCIENCIAS (Colombia); MSES (Croatia); RPF (Cyprus); Academy of
Sciences and NICPB (Estonia); Academy of Finland, ME, and HIP (Finland); CEA and
CNRS/IN2P3 (France); BMBF, DFG, and HGF (Germany); GSRT (Greece); OTKA and NKTH (Hungary);
DAE and DST (India); IPM (Iran); SFI (Ireland); INFN (Italy); NRF and WCU (Korea); LAS
(Lithuania); CINVESTAV, CONACYT, SEP, and UASLP-FAI (Mexico); PAEC (Pakistan); SCSR
(Poland); FCT (Portugal); JINR (Armenia, Belarus, Georgia, Ukraine, Uzbekistan); MST and
MAE (Russia); MSTD (Serbia); MICINN and CPAN (Spain); Swiss Funding Agencies
(Switzerland); NSC (Taipei); TUBITAK and TAEK (Turkey); STFC (United Kingdom); DOE and NSF
(USA). Individuals have received support from the Marie-Curie programme and the European
Research Council (European Union); the Leventis Foundation; the A. P. Sloan Foundation;
the Alexander von Humboldt Foundation; the Associazione per lo Sviluppo Scientifico e
Tecnologico del Piemonte (Italy); the Belgian Federal Science Policy Office; the Fonds
pour la Formation \`a la Recherche dans l'\'industrie et dans l'\'Agriculture
(FRIA-Belgium); and the Agentschap voor Innovatie door Wetenschap en Technologie
(IWT-Belgium).

\bibliography{auto_generated}   

\cleardoublepage\appendix\section{The CMS Collaboration \label{app:collab}}\begin{sloppypar}\hyphenpenalty=5000\widowpenalty=500\clubpenalty=5000\input{QCD-10-007-authorlist.tex}\end{sloppypar}
\end{document}

%% file: QCD-10-007-authorlist.tex
\textbf{Yerevan Physics Institute,  Yerevan,  Armenia}\\*[0pt]
V.~Khachatryan, A.M.~Sirunyan, A.~Tumasyan
\vskip\cmsinstskip
\textbf{Institut f\"{u}r Hochenergiephysik der OeAW,  Wien,  Austria}\\*[0pt]
W.~Adam, T.~Bergauer, M.~Dragicevic, J.~Er\"{o}, C.~Fabjan, M.~Friedl, R.~Fr\"{u}hwirth, V.M.~Ghete, J.~Hammer\cmsAuthorMark{1}, S.~H\"{a}nsel, C.~Hartl, M.~Hoch, N.~H\"{o}rmann, J.~Hrubec, M.~Jeitler, G.~Kasieczka, W.~Kiesenhofer, M.~Krammer, D.~Liko, I.~Mikulec, M.~Pernicka, H.~Rohringer, R.~Sch\"{o}fbeck, J.~Strauss, A.~Taurok, F.~Teischinger, P.~Wagner, W.~Waltenberger, G.~Walzel, E.~Widl, C.-E.~Wulz
% ...
\vskip\cmsinstskip
% ...
1:~~Also at CERN, European Organization for Nuclear Research, Geneva, Switzerland\\
% ...


%% file: QCD-10-007_temp.bbl
\providecommand{\href}[2]{#2}\begingroup\raggedright\begin{thebibliography}{10}%
\makeatletter
\providecommand{\hrefCMSnoop }[0]{\@secondoftwo}%
\makeatother

\bibitem{CMS_QCD-09-010}
\hrefCMSnoop {} {{ CMS} Collaboration, ``Transverse-momentum and pseudorapidity
  distributions of charged hadrons in $pp$ collisions at $\sqrt{s}$=0.9 and
  2.36 TeV'',} \textit{ JHEP} \textbf{ 02} (2010) 041,
  \href{http://www.arXiv.org/abs/1002.0621}{\texttt{ arXiv:1002.0621}}.
\href{http://dx.doi.org/10.1007/JHEP02(2010)041}{\texttt{
  doi:10.1007/JHEP02(2010)041}}.

\bibitem{CMS_QCD-10-006}
\hrefCMSnoop {} {{ CMS} Collaboration, ``{Transverse-momentum and
  pseudorapidity distributions of charged hadrons in $pp$ collisions at
  $\sqrt{s}$= 7 TeV}'',} \textit{ Phys. Rev. Lett.} \textbf{ 105} (2010)
  022002, \href{http://www.arXiv.org/abs/1005.3299}{\texttt{ arXiv:1005.3299}}.
\href{http://dx.doi.org/10.1103/PhysRevLett.105.022002}{\texttt{
  doi:10.1103/PhysRevLett.105.022002}}.

\bibitem{AliceStrange}
\hrefCMSnoop {} {{ ALICE} Collaboration, ``{Strange particle production in
  proton-proton collisions at $\sqrt{s}$ = 0.9 TeV with ALICE at the LHC}'',}
\href{http://www.arXiv.org/abs/1012.3257}{\texttt{ arXiv:1012.3257}}.

\bibitem{Aaij:2010nx}
\hrefCMSnoop {} {{ LHCb} Collaboration, ``{Prompt $K_S^0$ production in $pp$
  collisions at $\sqrt{s}=0.9$ TeV}'',} \textit{ Phys. Lett.} \textbf{ B693}
  (2010) 69, \href{http://www.arXiv.org/abs/1008.3105}{\texttt{
  arXiv:1008.3105}}.
\href{http://dx.doi.org/10.1016/j.physletb.2010.08.055}{\texttt{
  doi:10.1016/j.physletb.2010.08.055}}.

\bibitem{Pythia}
\hrefCMSnoop {} {{T. Sj\"ostrand, S. Mrenna and P. Skands}, ``PYTHIA 6.4
  physics and manual'',} \textit{ JHEP} \textbf{ 05} (2006) 026.
\href{http://dx.doi.org/10.1088/1126-6708/2006/05/026}{\texttt{
  doi:10.1088/1126-6708/2006/05/026}}.

\bibitem{Pop:2010qz}
\hrefCMSnoop {} {V.~T. Pop, M.~Gyulassy, J.~Barrette{ et~al.}, ``{Strong
  longitudinal color-field effects in $pp$ collisions at energies available at
  the CERN Large Hadron Collider}'',}
\href{http://www.arXiv.org/abs/1010.5439}{\texttt{ arXiv:1010.5439}}.

\bibitem{Koch:1986ud}
\hrefCMSnoop {} {P.~Koch, B.~{M\"uller}, and J.~Rafelski, ``{Strangeness in
  Relativistic Heavy Ion Collisions}'',} \textit{ Phys. Rept.} \textbf{ 142}
  (1986) 167.
\href{http://dx.doi.org/10.1016/0370-1573(86)90096-7}{\texttt{
  doi:10.1016/0370-1573(86)90096-7}}.

\bibitem{Abreu:2007kv}
\hrefCMSnoop {} {N.~Armesto, (ed.~) {et~al.}, ``{Heavy-ion collisions at the
  LHC -- Last Call for Predictions}'',} \textit{ J. Phys.} \textbf{ G35} (2008)
  054001, \href{http://www.arXiv.org/abs/0711.0974}{\texttt{ arXiv:0711.0974}}.
\href{http://dx.doi.org/10.1088/0954-3899/35/5/054001}{\texttt{
  doi:10.1088/0954-3899/35/5/054001}}.

\bibitem{Werner:2010zz}
\hrefCMSnoop {} {K.~Werner, I.~Karpenko, and T.~Pierog, ``{Collective effects
  in proton proton and heavy ion scattering, and the ``ridge'' at RHIC}'',}
  \textit{ J. Phys. Conf. Ser.} \textbf{ 230} (2010) 012026.
\href{http://dx.doi.org/10.1088/1742-6596/230/1/012026}{\texttt{
  doi:10.1088/1742-6596/230/1/012026}}.

\bibitem{Likhoded:2010pc}
\hrefCMSnoop {} {A.~K. Likhoded, A.~V. Luchinsky, and A.~A. Novoselov,
  ``{Inclusive light hadron production in $pp$ scattering at the LHC}'',}
  \textit{ Phys. Rev.} \textbf{ D82} (2010) 114006,
  \href{http://www.arXiv.org/abs/1005.1827}{\texttt{ arXiv:1005.1827}}.
\href{http://dx.doi.org/10.1103/PhysRevD.82.114006}{\texttt{
  doi:10.1103/PhysRevD.82.114006}}.

\bibitem{Chliapnikov:1988xa}
\hrefCMSnoop {} {P.~V. Chliapnikov, A.~K. Likhoded, and V.~A. Uvarov,
  ``{Inclusive Cross Sections in the Central Region and the Supercritical
  Pomeron}'',} \textit{ Phys. Lett.} \textbf{ B215} (1988) 417.
\href{http://dx.doi.org/10.1016/0370-2693(88)91458-X}{\texttt{
  doi:10.1016/0370-2693(88)91458-X}}.

\bibitem{CMS_Detector}
\hrefCMSnoop {} {{ CMS} Collaboration, ``The CMS experiment at the CERN LHC'',}
  \textit{ JINST} \textbf{ 3} (2008) S08004.
\href{http://dx.doi.org/10.1088/1748-0221/3/08/S08004}{\texttt{
  doi:10.1088/1748-0221/3/08/S08004}}.

\bibitem{Bartalini:2010su}
\hrefCMSnoop {} {P.~Bartalini, (ed.~) {et~al.}, ``{Proceedings of the First
  International Workshop on Multiple Partonic Interactions at the LHC
  (MPI08)}'',}
\href{http://www.arXiv.org/abs/1003.4220}{\texttt{ arXiv:1003.4220}}.

\bibitem{Geant4}
\hrefCMSnoop {} {{ GEANT4} Collaboration, ``{GEANT4 -- a simulation
  toolkit}'',} \textit{ Nucl. Instrum. Meth.} \textbf{ A506} (2003) 250.
\href{http://dx.doi.org/10.1016/S0168-9002(03)01368-8}{\texttt{
  doi:10.1016/S0168-9002(03)01368-8}}.

\bibitem{CMS_TRK-10-001}
\hrefCMSnoop {} {{ CMS} Collaboration, ``{CMS Tracking Performance Results from
  early LHC Operation}'',} \textit{ Eur. Phys. J.} \textbf{ C70} (2010) 1165,
  \href{http://www.arXiv.org/abs/1007.1988}{\texttt{ arXiv:1007.1988}}.
\href{http://dx.doi.org/10.1140/epjc/s10052-010-1491-3}{\texttt{
  doi:10.1140/epjc/s10052-010-1491-3}}.

\bibitem{PDG}
\hrefCMSnoop {} {{ Particle Data Group} Collaboration, ``{Review of particle
  physics}'',} \textit{ J. Phys.} \textbf{ G37} (2010) 075021.
\href{http://dx.doi.org/10.1088/0954-3899/37/7A/075021}{\texttt{
  doi:10.1088/0954-3899/37/7A/075021}}.

\bibitem{Skands:2009zm}
\hrefCMSnoop {} {P.~Z. Skands, ``{The Perugia Tunes}'',}
\href{http://www.arXiv.org/abs/0905.3418}{\texttt{ arXiv:0905.3418}}.

\bibitem{Pythia8}
\hrefCMSnoop {} {{T. Sj\"ostrand, S. Mrenna and P. Skands}, ``{A Brief
  Introduction to PYTHIA 8.1}'',} \textit{ Comput. Phys. Commun.} \textbf{ 178}
  (2008) 852, \href{http://www.arXiv.org/abs/0710.3820}{\texttt{
  arXiv:0710.3820}}.
\href{http://dx.doi.org/10.1016/j.cpc.2008.01.036}{\texttt{
  doi:10.1016/j.cpc.2008.01.036}}.

\bibitem{Alice_ppbar}
\hrefCMSnoop {} {{ ALICE} Collaboration, ``{Midrapidity antiproton-to-proton
  ratio in pp collisions at $\sqrt{s} = 0.9$ and $7$~TeV measured by the ALICE
  experiment}'',} \textit{ Phys. Rev. Lett.} \textbf{ 105} (2010) 072002,
  \href{http://www.arXiv.org/abs/1006.5432}{\texttt{ arXiv:1006.5432}}.
\href{http://dx.doi.org/10.1103/PhysRevLett.105.072002}{\texttt{
  doi:10.1103/PhysRevLett.105.072002}}.

\bibitem{Ansorge:1986xq}
\hrefCMSnoop {} {{ UA5} Collaboration, ``{Diffraction dissociation at the CERN
  pulsed $pp$ collider at c.m. Energies of 900 and 200 GeV}'',} \textit{ Z.
  Phys.} \textbf{ C33} (1986) 175.
\href{http://dx.doi.org/10.1007/BF01411134}{\texttt{ doi:10.1007/BF01411134}}.

\bibitem{Engel:1995sb}
\hrefCMSnoop {} {R.~Engel, J.~Ranft, and S.~Roesler, ``{Hard diffraction in
  hadron hadron interactions and in photoproduction}'',} \textit{ Phys. Rev.}
  \textbf{ D52} (1995) 1459,
  \href{http://www.arXiv.org/abs/hep-ph/9502319}{\texttt{
  arXiv:hep-ph/9502319}}.
\href{http://dx.doi.org/10.1103/PhysRevD.52.1459}{\texttt{
  doi:10.1103/PhysRevD.52.1459}}.

\bibitem{Bopp:1998rc}
\hrefCMSnoop {} {F.~W. Bopp, R.~Engel, and J.~Ranft, ``{Rapidity gaps and the
  PHOJET Monte Carlo}'',}
\href{http://www.arXiv.org/abs/hep-ph/9803437}{\texttt{ arXiv:hep-ph/9803437}}.

\bibitem{Tsallis}
\hrefCMSnoop {} {C.~Tsallis, ``{Possible Generalization of Boltzmann-Gibbs
  Statistics}'',} \textit{ J. Stat. Phys.} \textbf{ 52} (1988) 479.
\href{http://dx.doi.org/10.1007/BF01016429}{\texttt{ doi:10.1007/BF01016429}}.

\bibitem{Abelev:2006cs}
\hrefCMSnoop {} {{ STAR} Collaboration, ``Strange particle production in $p+p$
  collisions at $\sqrt{s}=200$ GeV'',} \textit{ Phys. Rev.} \textbf{ C75}
  (2007) 064901, \href{http://www.arXiv.org/abs/nucl-ex/0607033}{\texttt{
  arXiv:nucl-ex/0607033}}.
\href{http://dx.doi.org/10.1103/PhysRevC.75.064901}{\texttt{
  doi:10.1103/PhysRevC.75.064901}}.

\bibitem{Aaltonen:2011wz}
\hrefCMSnoop {} {{ CDF} Collaboration, ``{Production of $\Lambda^0$,
  $\overline{\Lambda}{}^0$, $\Xi^\pm$, and $\Omega^\pm$ Hyperons in $p\bar{p}$
  Collisions at $\sqrt{s} = 1.96$~TeV}'',}
\href{http://www.arXiv.org/abs/1101.2996}{\texttt{ arXiv:1101.2996}}.

\bibitem{Acosta:2005pk}
\hrefCMSnoop {} {{ CDF} Collaboration, ``{$K^0_S$ and $\Lambda^0$ production
  studies in $p\bar{p}$ collisions at $\sqrt{s}=$ 1800 and 630 GeV}'',}
  \textit{ Phys. Rev.} \textbf{ D72} (2005) 052001,
  \href{http://www.arXiv.org/abs/hep-ex/0504048}{\texttt{
  arXiv:hep-ex/0504048}}.
\href{http://dx.doi.org/10.1103/PhysRevD.72.052001}{\texttt{
  doi:10.1103/PhysRevD.72.052001}}.

\bibitem{Ansorge:1989ba}
\hrefCMSnoop {} {{ UA5} Collaboration, ``{Hyperon Production at 200 and 900 GeV
  C.M. Energy}'',} \textit{ Nucl. Phys.} \textbf{ B328} (1989) 36.
\href{http://dx.doi.org/10.1016/0550-3213(89)90090-4}{\texttt{
  doi:10.1016/0550-3213(89)90090-4}}.

\bibitem{Alner:1987wb}
\hrefCMSnoop {} {{ UA5} Collaboration, ``{UA5: A general study of
  proton-antiproton physics at $\sqrt{s}$ = 546 GeV}'',} \textit{ Phys. Rept.}
  \textbf{ 154} (1987) 247.
\href{http://dx.doi.org/10.1016/0370-1573(87)90130-X}{\texttt{
  doi:10.1016/0370-1573(87)90130-X}}.

\bibitem{Ansorge:1988fq}
\hrefCMSnoop {} {{ UA5} Collaboration, ``{Kaon production in $\bar{p}p$
  interactions at c.m. energies from 200 to 900 GeV}'',} \textit{ Z. Phys.}
  \textbf{ C41} (1988) 179.
\href{http://dx.doi.org/10.1007/BF01566915}{\texttt{ doi:10.1007/BF01566915}}.

\bibitem{Alner:1985ra}
\hrefCMSnoop {} {{ UA5} Collaboration, ``Kaon production in $\bar{p}p$
  reactions at a centre-of-mass energy of 540 GeV'',} \textit{ Nucl. Phys.}
  \textbf{ B258} (1985) 505.
\href{http://dx.doi.org/10.1016/0550-3213(85)90624-8}{\texttt{
  doi:10.1016/0550-3213(85)90624-8}}.

\bibitem{Alner:1984qa}
\hrefCMSnoop {} {{ UA5} Collaboration, ``{Observation of $\Xi^-$ Production in
  $p\bar{p}$ Interactions at 540 GeV CMS Energy}'',} \textit{ Phys. Lett.}
  \textbf{ B151} (1985) 309.
\href{http://dx.doi.org/10.1016/0370-2693(85)90859-7}{\texttt{
  doi:10.1016/0370-2693(85)90859-7}}.

\bibitem{Alexopoulos:1992ut}
\hrefCMSnoop {} {{ E735} Collaboration, ``{Hyperon production from $p\bar{p}$
  collisions at $\sqrt{s}$ = 1.8 TeV}'',} \textit{ Phys. Rev.} \textbf{ D46}
  (1992) 2773.
\href{http://dx.doi.org/10.1103/PhysRevD.46.2773}{\texttt{
  doi:10.1103/PhysRevD.46.2773}}.

\bibitem{Abe:1989hy}
\hrefCMSnoop {} {{ CDF} Collaboration, ``{$K_S^0$ production in $\bar{p}p$
  interactions at $\sqrt{s}=$ 630 GeV and 1800 GeV}'',} \textit{ Phys. Rev.}
  \textbf{ D40} (1989) 3791.
\href{http://dx.doi.org/10.1103/PhysRevD.40.3791}{\texttt{
  doi:10.1103/PhysRevD.40.3791}}.

\end{thebibliography}\endgroup
